%% file: wilsonlooppaper.tex
\documentclass[a4paper,twoside,10pt]{article}
\pdfoutput=1
\usepackage[USenglish]{babel} 
\usepackage[T1]{fontenc}
\usepackage[ansinew]{inputenc}
\usepackage{lmodern} 

\usepackage[numbers]{natbib}
\usepackage{ifthen}
\usepackage{graphicx} 

\usepackage{amsmath}
\usepackage{amsthm}
\usepackage{amsfonts}
\usepackage{amssymb}

\usepackage{a4wide} 

\usepackage[pdfusetitle,bookmarksnumbered,plainpages=false]{hyperref}

\input{commands}

\begin{document}

\begin{titlepage}

\title{Wilson loops in warped resolved deformed conifolds}
\begin{center}
{\Large \bf Wilson loops in warped resolved deformed conifolds}
\end{center}

\author{Stephen Bennett}
\begin{center}
Stephen Bennett \footnote{pystephen@swansea.ac.uk}
\end{center}

\begin{center}
\it{Department of Physics, University of Swansea \\
Singleton Park, Swansea SA2 8PP, United Kingdom}
\end{center}

\date{\today}
\begin{center}
\today
\end{center}

\begin{abstract}
We calculate quark-antiquark potentials using the relationship between the expectation value of the Wilson loop and the action of a probe string in the string dual. We review and categorize the possible forms of the dependence of the energy on the separation between the quarks. In particular, we examine the possibility of there being a minimum separation for probe strings which do not penetrate close to the origin of the bulk space, and derive a condition which determines whether this is the case. We then apply these considerations to the flavoured resolved deformed conifold background of Gaillard {\it et al} \citep{Gaillard:2010qg}. We suggest that the unusual behaviour we observe in this solution is likely to be related to the IR singularity which is not present in the unflavoured case.
\end{abstract}

\tableofcontents
\end{titlepage}

\setcounter{footnote}{0}

\pagestyle{plain} 

\input{introduction}
\input{eom}
\input{energystab}
\input{largeR}
\input{examples}
\input{NewSol}
		\input{rotated}
		\input{unrotated}
\input{discussion1}

\section*{Acknowledgments}
\addcontentsline{toc}{section}{Acknowledgments}
I would like to thank C. Nunez and G. Silva for valuable discussions. This work was supported by an STFC studentship.

\renewcommand\bibname{References}
\clearpage
\phantomsection
\addcontentsline{toc}{section}{References}
\bibliography{literature}


\end{document}

%% file: commands.tex
\providecommand{\abs}[1]{\left\lvert#1\right\rvert}

\def\d{\mathrm{d}}
\providecommand{\der}[2]{\frac{\d #1}{\d #2}}

\providecommand{\pder}[2]{\frac{\partial #1}{\partial #2}}

\def\tr{\mathop{\rm Tr}}

\providecommand{\h}[1]{{\hat{#1}}}

\newcommand{\nn}{{\nonumber}}

\newcommand{\cC}{{\cal C}}
\newcommand{\cD}{{\cal D}}

\newcommand{\cF}{{\cal F}}

\newcommand{\cN}{{\cal N}}
\newcommand{\cR}{{\cal R}}
\newcommand{\cO}{{\cal O}}
\newcommand{\cT}{{\cal T}}

\newcommand{\q}{\theta}

\newcommand{\vph}{\varphi}
\newcommand{\w}{\omega}

\newcommand{\Nc}{N_\text{c}}
\newcommand{\Nf}{N_\text{f}}

\newcommand{\SNG}{S_\text{NG}}
\newcommand{\Teff}{T_\text{eff}}
\newcommand{\TDt}{T_\text{D3}}
\newcommand{\TDo}{T_\text{D1}}
\newcommand{\Tst}{T_\text{str}}

\newcommand{\dsint}{\d s_{\text{int}}^2}
\newcommand{\dseff}{\d s_{\text{eff}}^2}
\newcommand{\Linf}{L_\infty}
\newcommand{\Einf}{E_\infty}
\newcommand{\ex}{\Delta}

\newcommand{\Phin}{\Phi_\infty}
\newcommand{\Phio}{\Phi_0}

\newcommand{\Rb}{\h R_0}		
\newcommand{\Rmin}{R_0}			
\newcommand{\Rinf}{R_\infty}

\newcommand{\Tb}{\h T_0}		
\newcommand{\Tmin}{T_0}			

\newcommand{\rhb}{\h\rho_0}	
\newcommand{\rhmin}{\rho_0}	
\newcommand{\rhoM}{\rho_1}	

\newcommand{\rb}{\h r_0}	
\newcommand{\rmin}{r_0}		
\newcommand{\rinf}{r_\infty}

\newcommand{\qt}{\tilde{\theta}}
\newcommand{\vpht}{\tilde{\varphi}}
\newcommand{\wt}{\tilde{\omega}}

\newcommand{\Po}{P}

\newcommand{\hyF}{\;{}_2F_1\!}

%% file: introduction.tex
\section{Introduction}\label{sec:intro}
One of the most important observables in a field theory is the Wilson loop \citep{PhysRevD.10.2445}, given by
\begin{align}
	W(\cC) = \frac{1}{\Nc}\tr \Po e^{ i \oint_\cC A },
\end{align}
for a closed curve $\cC$. In particular, the potential $E$ of a quark-antiquark pair is related to the VEV of a Wilson loop. Indeed, for a separation $L$, take the loop $\cC$ to be a rectangle of sides $L$, $\cT$, with $\cT\to\infty$, then
\begin{align}
	\left\langle W(\cC) \right\rangle	\sim	e^{-E \cT}.  \label{eq:WLEnergy}
\end{align}

\begin{figure}[htbp]
	\centering
		\includegraphics{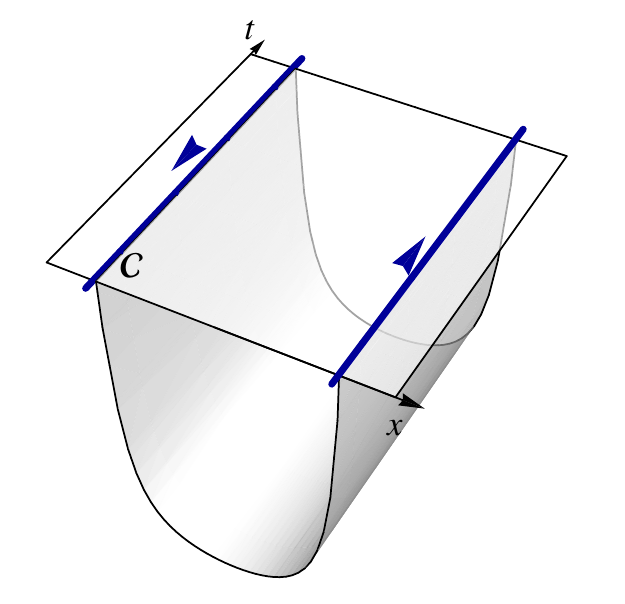}
	\caption{Schematic diagram of how the Wilson loop $\cC$ (blue) relates to the world-sheet of a string extending into the bulk. The loop should be taken to extend an infinite distance in the $t$ direction.}
	\label{fig:wilsonLoop}
\end{figure}

The Wilson loop is particularly significant in the context of gauge-string duality and the Maldacena conjecture \citep{Maldacena:1997re}, because it is accessible from the string side of the correspondence. For a review see \citep{Sonnenschein:1999if}. As proposed in \citep{Rey:1998ik,Maldacena:1998im}, the associated quantity in the dual string theory is the action of a string world-sheet which ends on $\cC$ at the boundary of the AdS space (\autoref{fig:wilsonLoop}). That is
\begin{align}
	W(\cC) = \int_{\partial F(\cC)}\cD F e^{-S[F]},
\end{align}
where $F$ describes the fields of the string theory, with boundary values $\partial F(\cC)$. In the limit of strong coupling the result is that the Wilson loop corresponds to the area of a surface bounded by $\cC$, extending into the bulk and forming the world-sheet of a classical string. This means that in the strong coupling regime of the QFT
\begin{align}
	\left\langle W(\cC) \right\rangle \sim e^{-\SNG},
\end{align}
and referring to \autoref{eq:WLEnergy}, the energy of the quark-antiquark pair corresponds to the Nambu-Goto action of the string \citep{Sonnenschein:1999if,Nunez:2009da}
\begin{align}
	E=\frac{\SNG}{\cT}.
\end{align}
This is divergent, and is renormalised by subtracting the (infinite) quark masses, given by the action of two rods from the ends of the string to the end of the space, as descibed in \citep{Maldacena:1998im,Sonnenschein:1999if,Nunez:2009da}. This is discussed more rigourously in \citep{Drukker:1999zq}.

The aim is therefore to solve the equations of motion resulting from the action and so determine the shape of the string formed for a given $L$ in a given background. In \autoref{sec:actionEOM} we discuss the equations of motion for a string, following the derivation in \citep{Sonnenschein:1999if,Nunez:2009da}, and describe a generalisation to D branes. We discuss the possible behaviour of the function $E(L)$, and of the string shape in sections \ref{sec:energystab}-\ref{sec:largeR}. We then demonstrate the results of the preceding sections with respect to some well-understood backgrounds (\autoref{sec:examples}), before applying them to the flavoured resolved deformed conifold \citep{Gaillard:2010qg} in \autoref{sec:NewSol}.

%% file: eom.tex
\section{The action and equations of motion	}\label{sec:actionEOM}
\subsection{Action}
We will consider backgrounds of the form
\begin{align}
	ds^2	= g_{\mu\nu}\d x^\mu \d x^\nu = -	g_{tt}\d t^2	+	g_{xx}\d\vec{x}^2	+	g_{\rho\rho}\d\rho^2	+	g_{ij}\d\q^i\d\q^j,		\label{eq:genMetric}
\end{align}
where $g_{tt}$, $g_{xx}$ and $g_{\rho\rho}$ are functions only of $\rho$. Here and throughout this paper we use the string frame. We will restrict our attention to $p$-dimensional objects which extend on time and one spatial Minkowski direction $x$, and probe the radial direction according to
\begin{align}
	x=x(X^1),	\qquad\qquad	\rho=\rho(X^1),		\label{eq:embedding}
\end{align}
where we use world-volume coordinates $X^\alpha$, $0\le \alpha\le p$. For $p>1$, the object also extends in the internal space described by the coordinates $\q^i$. We are interested only in the static case, so we can identify $X^0=t$. 

The  action for such an object, with tension $T_0$, is 
\begin{align}
	S = T_0 \int \d^{p+1}X e^{-\alpha \Phi}\sqrt{-\det G_{\alpha\beta}},		\label{eq:genAction}
\end{align}
where 
\begin{align}
	G_{\alpha\beta}	=	g_{\mu\nu} \pder{x^\mu}{X^\alpha}\pder{{x^\nu}}{{X^\beta}}
\end{align}
is the induced metric on the world volume, and $\alpha=1$ for a D$p$ brane, with $\alpha=0$ otherwise. For the configuration described, the induced metric is
\begin{align}
	ds_\text{induced}^2 &= -	g_{tt} (\d X^0)^2
												+ \left(	g_{xx} {x'}^2  + g_{\rho\rho} {\rho'}^2
													\right) (\d X^1)^2
												+ G_{ab}^{(p-1)}\d X^a \d X^b,												\\
		G_{ab}^{(p-1)}		&=	g_{ij}\pder{{\q^i}}{{X^a}}\pder{{\q^j}}{{X^b}},
\end{align}
where $x'=\d x/\d X^1$. Writing the time interval as $\cT$, the action \eqref{eq:genAction} is then
\begin{align}
	S = T_0 \cT \int \d^pX\; e^{-\alpha\Phi} \sqrt{	g_{tt}
																											\left(	g_{xx} {x'}^2  + g_{\rho\rho} {\rho'}^2		\right)
																											\det G_{ab}^{(p-1)}
																										}.
\end{align}
This can be written in a form corresponding to a 1-dimensional `effective string' with tension $\Tst(\rho)$. Defining $f(\rho)^2=g_{tt}g_{xx}$ and $g(\rho)^2=g_{tt}g_{\rho\rho}$,
\begin{align}
	S = \cT \int \d X^1\;  \left( f\sqrt{	{x'}^2 + \frac{g^2}{f^2} {\rho'}^2	} \right)
												\;	\Tst,																								\label{eq:1dAction}
\end{align}
where
\begin{align}
 \Tst(\rho) = T_0 e^{-\alpha\Phi} \int \d^{p-1}X\; \sqrt{\det G_{ab}^{(p-1)}}.
\end{align}
This has a simple interpretation: The tension $\Tst$ is as expected the energy density on the effective string, while the factor
\begin{align}
	\d X^1\;  f\sqrt{	{x'}^2 + \frac{g^2}{f^2} {\rho'}^2	}	\label{eq:lengthel}
\end{align}
is the length element on a string embedded according to \eqref{eq:embedding} in the geometry \eqref{eq:genMetric}. The action \eqref{eq:1dAction} is therefore an obvious generalisation of the case considered in \citep{Sonnenschein:1999if,Nunez:2009da} to a string with $\rho$-dependent tension. In fact \eqref{eq:1dAction} can also be obtained from the action used in \citep[section II.A]{Nunez:2009da} by the replacements
\begin{align}
	f\to \Tst f, \qquad\qquad g\to \Tst g.		\label{eq:genReplace}
\end{align}
\begin{figure}[htbp]
	\centering
		\includegraphics{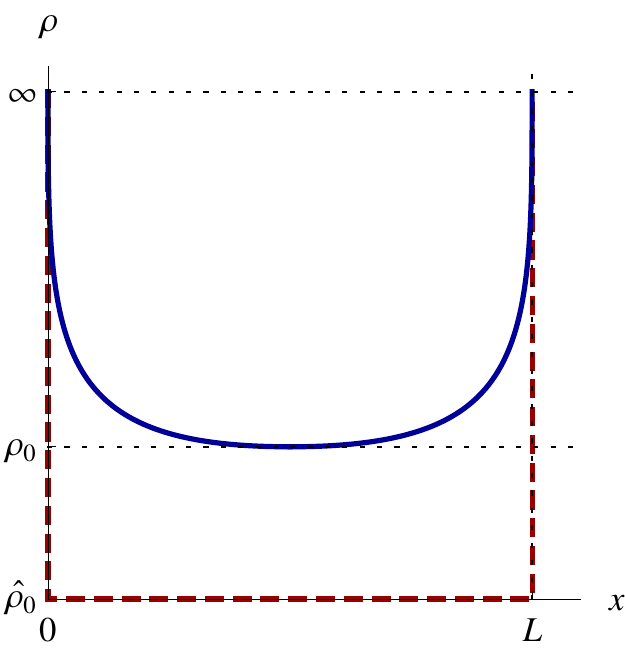}
	\caption{The generic shape of the effective string (solid line). The dashed line shows the `free' solution \eqref{eq:zeroSol}, discussed in \autoref{sec:eom}.}
	\label{fig:genericstring}
\end{figure}

\subsection{Rescaled radial coordinate}\label{sec:rescaleToR}
The following discussion will be considerably simplified by the introduction of a rescaled coordinate $R$ defined by
\begin{align}
	\der{R}{\rho}	=	\frac{g}{f}.
\end{align}
Then the metric \eqref{eq:genMetric} becomes
\begin{align}
	ds^2 = -g_{tt} \d t^2 + \frac{f^2}{g_{tt}} \left( \d\vec{x}^2 + \d R^2 \right) + g_{ij} \d\q^i\d\q^j,
	\label{eq:confMetric}
\end{align}
and the action \eqref{eq:1dAction} is
\begin{align}
	S = \cT \int \d X^1\;  \sqrt{	{x'}^2 +  {R'}^2	}
												\;	\Teff,											\label{eq:RAction}
\end{align}
where $\Teff=f\Tst$. Notice we can interpret this action as being that of a string in any metric of the form
\begin{align}
	\dseff = -\gamma_{tt} \d t^2 + \gamma_{xx} (\d x^2 + \d R^2),
\end{align}
provided we give the string a tension equal to ${\Teff}/{\sqrt{\gamma_{tt}\gamma_{xx}}}$. In particular, interpreting $\Teff$ itself as the string tension results in Minkowski space. The behaviour of the effective string is then described completely by the function $\Teff$. This is in contrast to the description in terms of the original metric in equations \eqref{eq:genMetric} and \eqref{eq:confMetric}, for which the natural interpretation involved a geometric factor \eqref{eq:lengthel} as well as an $R$-dependent tension $\Tst$.

Although useful in the general discussion, the integration involved in obtaining $R(\rho)$ means that this coordinate system will be difficult to apply to any specific case except for extremely simple backgrounds.

\subsection{Equations of motion and separation}\label{sec:eom}
The derivation of the equation of motion from \eqref{eq:RAction} is essentially the same as the calculation in \citep{Sonnenschein:1999if,Nunez:2009da}. Imposing time independence, we get the single equation
\begin{align}
	C\der{R}{X^1} = \pm \der{x}{X^1} \sqrt{\Teff^2 - C^2}.		\label{eq:EoM}
\end{align}

Parametrising the effective string as 
\begin{align}
	X^1=x,	\qquad\qquad R=R(x),		\label{eq:paramR(x)}
\end{align}
we can integrate to obtain the shape of a string with a minimum radial coordinate at $R=\Rmin$. We impose $C=\Teff(\Rmin)$, and obtain
\begin{align}
	x(R) = \begin{cases}
					\displaystyle		\int_R^{\Rinf}\d R' \frac{ \Teff(\Rmin) }{\sqrt{\Teff(R')^2 - \Teff(\Rmin)^2  }}	,	\qquad
					&\displaystyle	0\le x \le \frac{L}{2}	,		\\
					\displaystyle		L-\int_R^{\Rinf}\d R' \frac{ \Teff(\Rmin)}{\sqrt{\Teff(R')^2 - \Teff(\Rmin)^2  }},	\qquad
					&\displaystyle	\frac{L}{2} \le x \le L,
					\end{cases}		\label{eq:smoothSol}
\end{align}
where
\begin{align}
	L(\Rmin) = 2\int_{\Rmin}^{\Rinf}\d R \frac{ \Teff(\Rmin) }{\sqrt{\Teff(R)^2 - \Teff(\Rmin)^2  }}		\label{eq:L(R)}
\end{align}
is the separation of the endpoints of the effective string and $\Rinf=R(\rho\to\infty)$. This is the same as is obtained by modifying the result of \citep{Nunez:2009da} according to the prescription \eqref{eq:genReplace}.

In some cases we will find that $T(\Rb)=0$, where $\Rb$ is the minimum radial coordinate contained in the space. Then there is an additional solution to \eqref{eq:EoM} (with $C=0$) which is not compatible with the parametrisation \eqref{eq:paramR(x)}. This corresponds to a string  which drops vertically from the endpoints and stretches horizontally along the `bottom of the space', $R=\Rb$, as shown in \autoref{fig:genericstring}. A suitable parametrisation is
\begin{align}
	(x,R) = \begin{cases}
								(0,\  \Rb-X^1),	\qquad		&	X^1\le0,		\\
								(X^1,\ \Rb),		\qquad		&	0\le X^1 \le L,	\\
								(L,\ \Rb-L+X^1),	\qquad		& X^1 \ge L.
					\end{cases}																								\label{eq:zeroSol}
\end{align}

As we shall see, generically $L(\Rmin)$ has inversion points, and together with the possibility of the extra solution \eqref{eq:zeroSol} this means that $\Rmin(L)$ can be multivalued. The different branches can be interpreted as corresponding to stable, metastable and unstable configurations for the effective string (see \autoref{sec:energystab}).

\subsection{Boundary conditions in the UV}\label{sec:UVBCs}
When we consider a fundamental string we must enforce Dirichlet boundary conditions at $R\to\Rinf$, as described in \citep{Nunez:2009da}. This corresponds to the string ending on a D-brane at large $R$. Specifically, we require that
\begin{align}
	\der{x}{X^1}\to0
\end{align}
for $R\to\Rinf$. Referring to \eqref{eq:EoM} this means that we need
\begin{align}
	\lim_{R\to\Rinf} \frac{\Teff(R)^2-C^2}{C^2}	=	\infty.
\end{align}
Recalling that we have imposed $C=\Teff(\Rmin)$ and that for a fundamental string $\Teff=f$, this becomes
\begin{align}
		\lim_{R\to\Rinf} \frac{f(R)^2-f(\Rmin)^2}{f(\Rmin)^2}	=	\infty.
\end{align}
As this must hold for all $\Rmin$, we can simply require that
\begin{align}
		\lim_{R\to\Rinf} f(R)	=	\infty.	\label{eq:UVBCs}
\end{align}

Although this condition is not required when the string is replaced by a D-brane, we will restrict our attention to those backgrounds in which \eqref{eq:UVBCs} holds.

%% file: energystab.tex
\section{Energy and stability}\label{sec:energystab}
As noted in \autoref{sec:intro}, the energy of the quark-antiquark pair is simply given by $S/\cT$, which as would be expected corresponds to the tension integrated along the string. This is in general infinite, and so we renormalise by subtracting the action of two vertical rods extending from $\Rb$ to infinity \citep{Maldacena:1998im,Sonnenschein:1999if,Drukker:1999zq,Nunez:2009da}. For the smooth solution \eqref{eq:smoothSol}, which we can parametrise as $R(x)$, this gives
\begin{align}
	E(\Rmin) =	2\int_{\Rmin}^{\Rinf} \d R \sqrt{1+\left(\der{x}{R}\right)^2} \;\Teff - 2\int_{\Rb}^{\Rinf} \d R \;\Teff.
\end{align}
Using the equation of motion \eqref{eq:EoM}, this can be written
\begin{align}
	E(\Rmin) = 2\int_{\Rmin}^{\Rinf} \d R\ \Teff(R)
																				 \left[ \frac{\Teff(R)}
																				 						 {\sqrt{\Teff(R)^2 - \Teff(\Rmin)^2 }}
																				 				-1
																				 \right]
							-	2\int_{\Rb}^{\Rmin} \d R \;\Teff(R).		\label{eq:energyR}
\end{align}

Given a form for the function $L(\Rmin)$ it is simple to obtain the qualitative behaviour of $E(L)$ without evaluating \eqref{eq:energyR}. Generalising the result obtained in \citep{Nunez:2009da,Brandhuber:1999jr} using \eqref{eq:genReplace} we have that the force is
\begin{align}
	\der{E}{L} = \Teff(\Rmin).		\label{eq:dEdL}
\end{align}
Given the reasonable assumption $\Teff(\Rmin)$ is continuous and positive this implies that the extrema of $E(\Rmin)$ correspond to those of $L(\Rmin)$, and to cusps in $E(L)$.

\begin{figure}[htbp]
	\centering
		\includegraphics{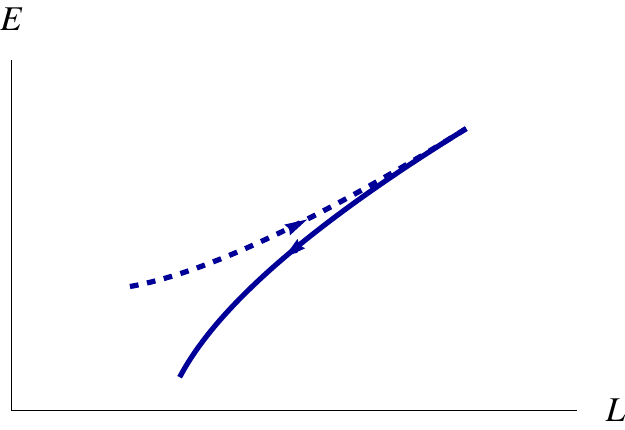}
	\caption{A generic cusp in $E(L)$. The concavity condition \citep{PhysRevD.33.2723} leads us to expect that the upper (dotted) branch is unstable. The arrows show the direction of increasing $\Rmin$.}
	\label{fig:genericCusp}
\end{figure}

The possible presence of extrema in the function $L(\Rmin)$ raises the question of which branches of the solution represent physical configurations of the string for a given $L$. It was shown in \citep{Avramis:2006nv} that these extrema also correspond to the boundaries between stable and unstable configurations (ignoring any regions of instability due to fluctuations in the angular directions $\q^i$). Although this implies that only one side of the extremum describes a physical solution (as opposed to one being stable and the other metastable), it is not clear how to identify the physical branch. However, we can make progress if we assume that $\Teff(\Rmin)$ is always increasing with $\Rmin$. Referring to \eqref{eq:dEdL}, this is equivalent to the statement that
\begin{align}
	\der{}{\Rmin}\left[ E'\bigl(L(\Rmin)\bigr) \right] >0.
\end{align}
In terms of the function $E(L)$, we see that $E''(L)$ changes sign at each cusp (\autoref{fig:genericCusp}), with $E''(L)>0$ for the upper (higher $E$) branch and $E''(L)<0$ for the lower. We can relate this to the concavity condition discussed in \citep{PhysRevD.33.2723}, namely that for a physical interaction between quarks we must have $E''(L)\le0$. We therefore expect that the upper branch at each cusp is unstable.

\begin{figure}[htbp]
	\centering
		\includegraphics{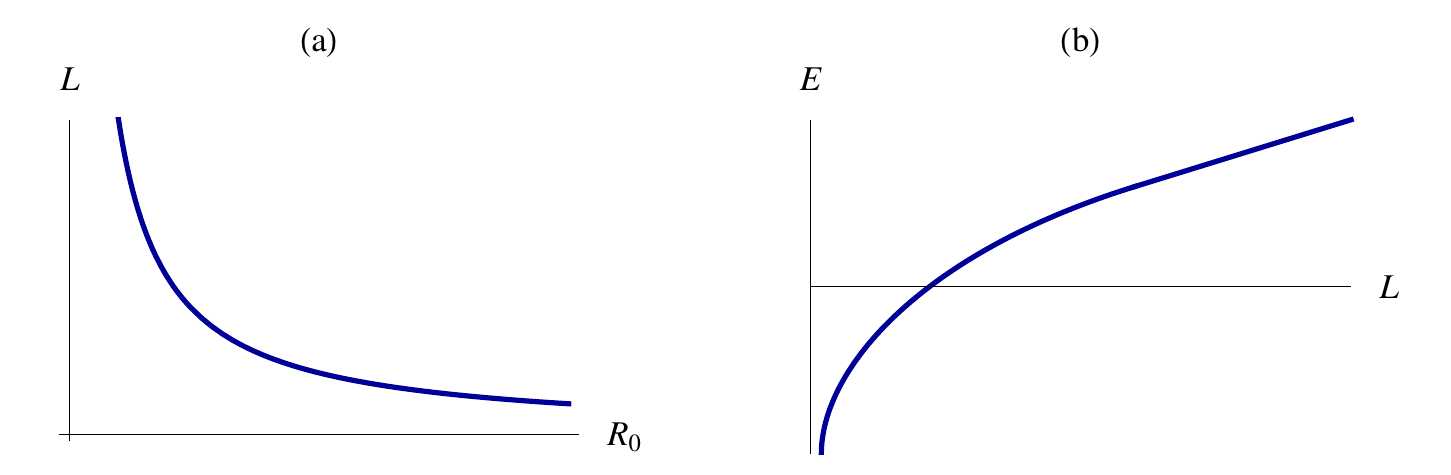}
	\caption{The qualitative behaviour of (a) $L(\Rmin)$ and (b) $E(L)$ in a simple confining case, such as Klebanov-Strassler.}
	\label{fig:confining}
\end{figure}

Probably the simplest form of behaviour occurs when $\Teff(\Rb)\ne0$. By the argument of \citep{Kinar:1998vq,Sonnenschein:1999if} this means that $E(L)$ becomes linear at large $L$, corresponding to confinement. If $L(\Rmin)$ is decreasing for all $\Rmin$, we obtain the qualitatively simple behaviour exhibited, for example, by the Klebanov-Strassler \citep{Klebanov:2000hb} and wrapped D5 brane \citep{Maldacena:2000yy} models. This is depicted schematically in \autoref{fig:confining}.

\begin{figure}[htbp]
	\centering
		\includegraphics{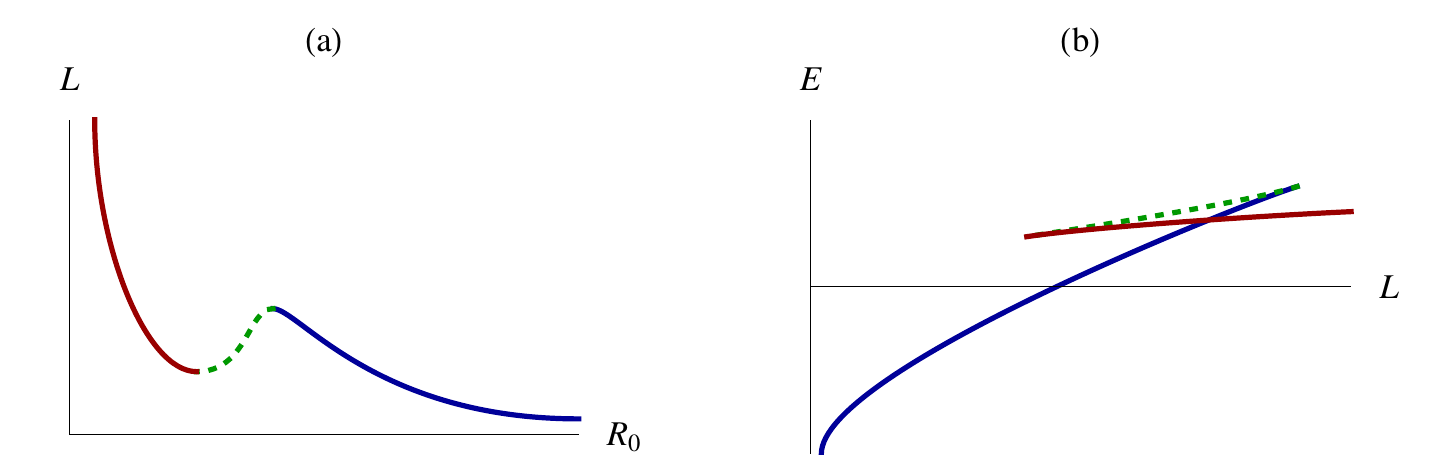}
	\caption{The qualitative behaviour of (a) $L(\Rmin)$ and (b) $E(L)$ in the `van der Waals' case. The dotted region is unphysical and is expected to correspond to an unstable string configuration.}
	\label{fig:vanderWaals}
\end{figure}

More complicated behaviour is exhibited in the models with massive dynamical flavours \citep{Bigazzi:2008gd,Bigazzi:2008zt,Bigazzi:2008cc,Bigazzi:2008qq}, and also the walking D5 background discussed in detail in \citep{Nunez:2008wi,Nunez:2009da}. There $L(\Rmin)$ has two local extrema, leading to two cusps in $E(L)$, as shown in \autoref{fig:vanderWaals}. We still have $T(\Rb)\ne0$, and confinement is again seen for large $L$. In \citep{Nunez:2009da} an analogy with a van der Waals gas was proposed. This again suggests that the upper branch at the cusps, corresponding to $L'(\Rmin)>0$, should be identified as unstable.

When $\Teff(\Rb)=0$ we also obtain the second solution \eqref{eq:zeroSol}. As we renormalise by subtracting the action of two vertical rods, and there is no contribution to the energy from the part of the string with $R=\Rb$, this solution has $E=0$ independent of $L$. This is the stable solution for sufficiently large $L$, so at large separations the endpoints of the string behave like free particles.

As pointed out in \citep{Avramis:2006nv}, this is analogous to the case of a soap film stretched between two circular rings. In fact, if $\Teff(R)\propto R$, corresponding to a string in Rindler space, the analogy becomes exact \citep{Avramis:2007mv}, as the action \eqref{eq:RAction} is then identical to that of the soap film. The `free' solution \eqref{eq:zeroSol} corresponds in the case of the soap film to a disconnected configuration, with the film forming a disc over each of the rings. 

\begin{figure}[htbp]
	\centering
		\includegraphics{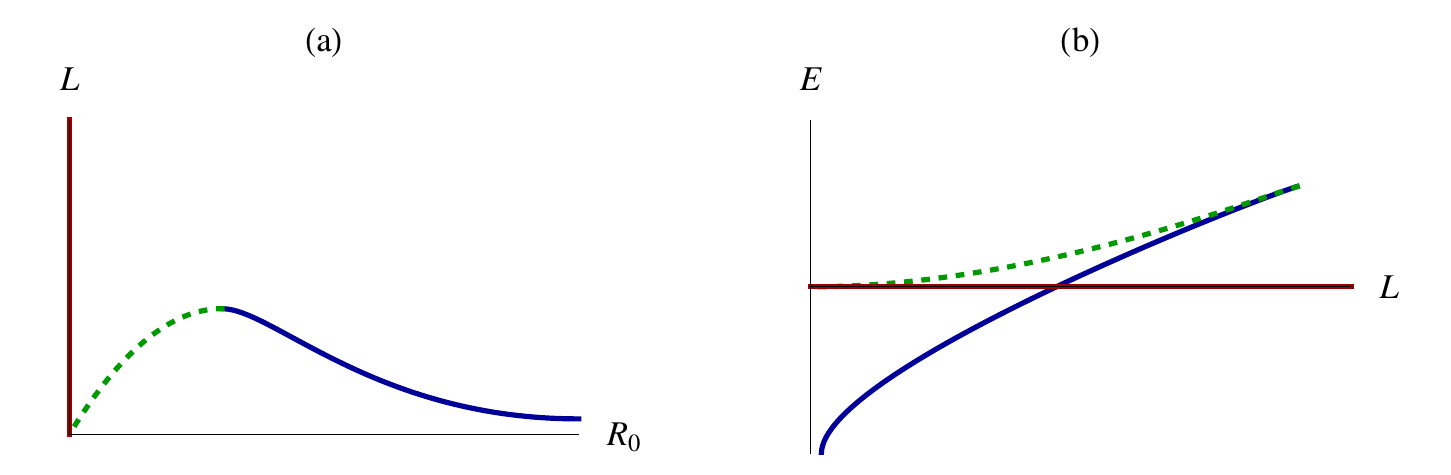}
	\caption{The qualitative behaviour of (a) $L(\Rmin)$ and (b) $E(L)$ in the `soap film' case. The red region now corresponds to the solution \eqref{eq:zeroSol}. }
	\label{fig:soapfilm}
\end{figure}

Additionally, when $\Rmin=\Rb$ the integrand in \eqref{eq:L(R)} vanishes for all $R\ne\Rmin$. This means that unless the lower limit of the integral gives a non-zero contribution the separation given by the smooth solution \eqref{eq:smoothSol} will go to zero as the string approaches the end of the space; $L(\Rb)=0$. As can be seen from \autoref{fig:soapfilm}, this can be considered a special case of the `van der Waals' behaviour (\autoref{fig:vanderWaals}), in which the minimum in $L(\Rmin)$ moves to the origin.

\begin{figure}[htbp]
	\centering
		\includegraphics{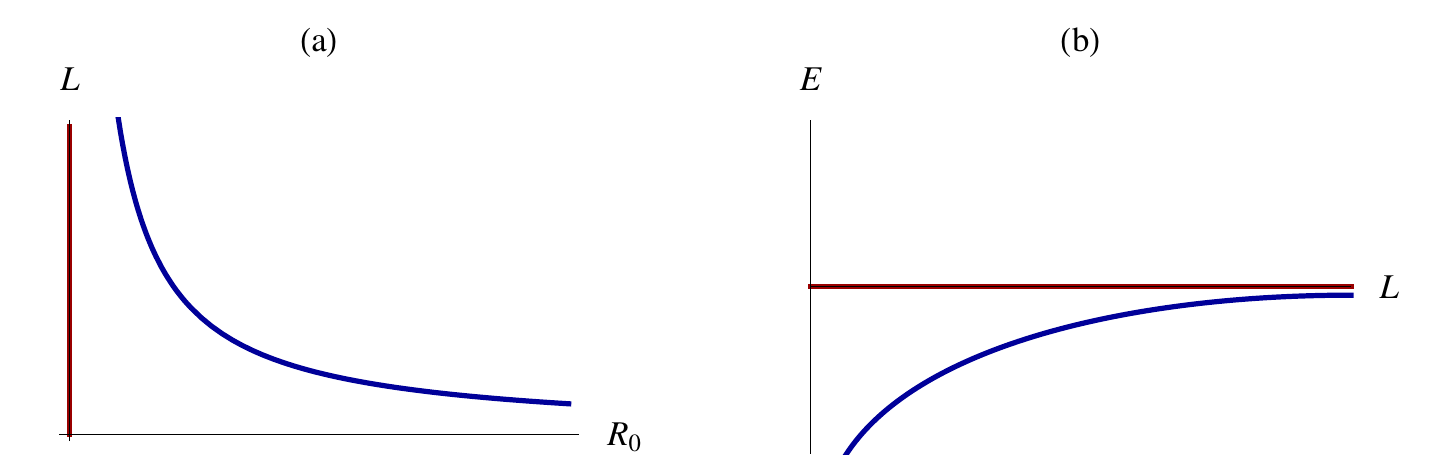}
	\caption{The qualitative behaviour of (a) $L(\Rmin)$ and (b) $E(L)$ in the `Coulomb' case. }
	\label{fig:coulomb}
\end{figure}

If the lower limit of the integral in \eqref{eq:L(R)} diverges, the separation will diverge for $\Rmin\to\Rb$ despite having $\Teff(\Rb)=0$. This is the case for a string in $AdS_5\times S^5$. As before, we can consider this a special case of the `soap film' case, with the maximum in $L(\Rmin)$ moving to $(\Rmin=\Rb,L=\infty)$, as shown in \autoref{fig:coulomb}. This results in a qualitatively Coulombic potential (exact in the case of $AdS$.). The `free' solution is now presumably metastable for all $L$.

Finally, it is possible for any of these forms of behaviour to be further modified by the development of an additional local maximum and minimum, as in the `van der Waals' case, \autoref{fig:vanderWaals}. This will again result in a pair of cusps.

%% file: largeR.tex
\section{General results on the behaviour of $L$ and $E$ for large $\Rmin$}\label{sec:largeR}
In the discussion of \autoref{sec:energystab} it was implicitly assumed that $L\to0$ for $\Rmin\to\Rinf$. This is not always the case. We will find cases in which instead we have $L\to\Linf\ne0$.

Assuming that $ T(\Rmin)\to\infty$ for $\Rmin\to\Rinf$, we can write the separation \eqref{eq:L(R)} as
\begin{align}
	L(\Rmin)	=	2	\int_{\Tmin}^\infty	\d T\; \frac{ T}{ T'(R)}
																											\frac{1}{ T\sqrt{	 T^2/\Tmin^2 - 1		}}, \label{eq:LintdT}
\end{align}
where in this section we write $T\equiv\Teff(R)$ and $\Tmin\equiv\Teff(\Rmin)$. Defining $t=T/\Tmin$, this is simply
\begin{align}
		L(\Rmin)	=	2	\int_1^\infty	\d t\; \frac{t}{t'(R)}
																			 \frac{1}{t\sqrt{	t^2 - 1		}},	\label{eq:Lintdt}
\end{align}
in which the $\Rmin$ dependence is entirely contained in the factor
\begin{align}
	\frac{t}{t'(R)}	=	\frac{ T(R(t,\Rmin))}{ T'(R(t,\Rmin))},		\label{eq:LR0dep}
\end{align}
where the factors of $\Tmin$ from the definition of $t$ cancel.

We are interested in the case when the separation is constant for large $\Rmin$. The integrand in \eqref{eq:Lintdt} must therefore be a function only of $t$, which is equivalent to requiring that \eqref{eq:LR0dep} is a function only of $t$ for large $R$. That is
\begin{align}
	\frac{T}{ T'(R)}	=	\cF\left( \frac{T}{\Tmin} \right)		\qquad\qquad \text{for $R\to\Rinf$}.
\end{align}
However, the left hand side is explicitly independent of $\Rmin$ so this can only be satisfied if $\cF(t)$ is a constant. We therefore have $L(\Rmin\to\Rinf)=\text{constant}$ if and only if
\begin{align}
	\frac{T}{ T'(R)}	=	\text{constant}		\qquad\qquad \text{for $R\to\Rinf$}.	\label{eq:condForConst}
\end{align}
In this case the integral \eqref{eq:Lintdt} can be evaluated, giving
\begin{align}
	\Linf \equiv \lim_{\Rmin\to\Rinf} L(\Rmin)	=	\pi \lim_{R\to\Rinf} \left(\frac{T}{ T'(R)}\right).	\label{eq:LinfR}
\end{align}



The above calculation of $\Linf$ relied on the fact that the integral for $L(\Rmin)$ covers only the range $\Rmin\le R<\Rinf$. The integral \eqref{eq:energyR} for $E(\Rmin)$ covers the whole range $\Rb<R<\Rinf$, and so an analogous calculation is not possible. When $L(\Rmin\to\Rinf)\to\text{constant}$, it is however possible to find a condition which determines whether $E(\Rmin\to\Rinf)\to\text{constant}$.

Referring to \eqref{eq:LinfR}, for large $R$ we can write 
\begin{align}
	\frac{T}{T'(R)} = \frac{\Linf}{\pi}	+	\ex'(T(R)),	\label{eq:defexprime}
\end{align}
where $\ex'(T(R))\to0$ for $R\to\Rinf$, so that by \eqref{eq:Lintdt}
\begin{align}
		L(\Rmin)	=	\Linf	+
								2	\int_1^\infty	\d t\; \ex'(T(t,\Rmin))
																			 \frac{1}{t\sqrt{	t^2 - 1		}}.
\end{align}
The energy is given by \eqref{eq:energyR}, which becomes
\begin{align}
	E(\Rmin)	=	2	\int_{\Tmin}^\infty \d T\; \left[\frac{\Linf}{\pi} + \ex'(T) \right] \left(\frac{T}{\sqrt{T^2-\Tmin^2}} -1 \right)
							- 2 \int_{\Tb}^\infty \d T\; \left[\frac{\Linf}{\pi} + \ex'(T) \right],
\end{align}
where $\Tb\equiv\Teff(\Rb)$. After evaluating some integrals this can be written as
\begin{align}
	E(\Rmin)	=	\frac{2\Linf\Tb}{\pi} + 2\ex(\Tb) - 2\ex(\Tmin)
							+ 2 \int_{\Tmin}^\infty \d T\; \ex'(T) \left( \frac{T}{\sqrt{T^2-\Tmin^2}}  -1 \right).
	\label{eq:E(ex)}
\end{align}
The first two terms are constants, and the last two, which contain the $\Rmin$ dependence, involve only the region $\Rmin\le R<\Rinf$. Whether the energy approaches a constant is therefore determined by the large $R$ behaviour of $\ex'(T(R))$. The condition is that $E(\Rmin)\to\text{constant}$ for $\Rmin\to\Rinf$ if and only if $\ex'(T)$ vanishes at least as fast as
\begin{align}
	\ex'(T)\sim \frac{1}{T^{1+\epsilon}},	\qquad\qquad	\epsilon>0		\label{eq:condexprime}
\end{align}
for $T\to\infty$.

\begin{figure}[htbp]
	\centering
		\includegraphics{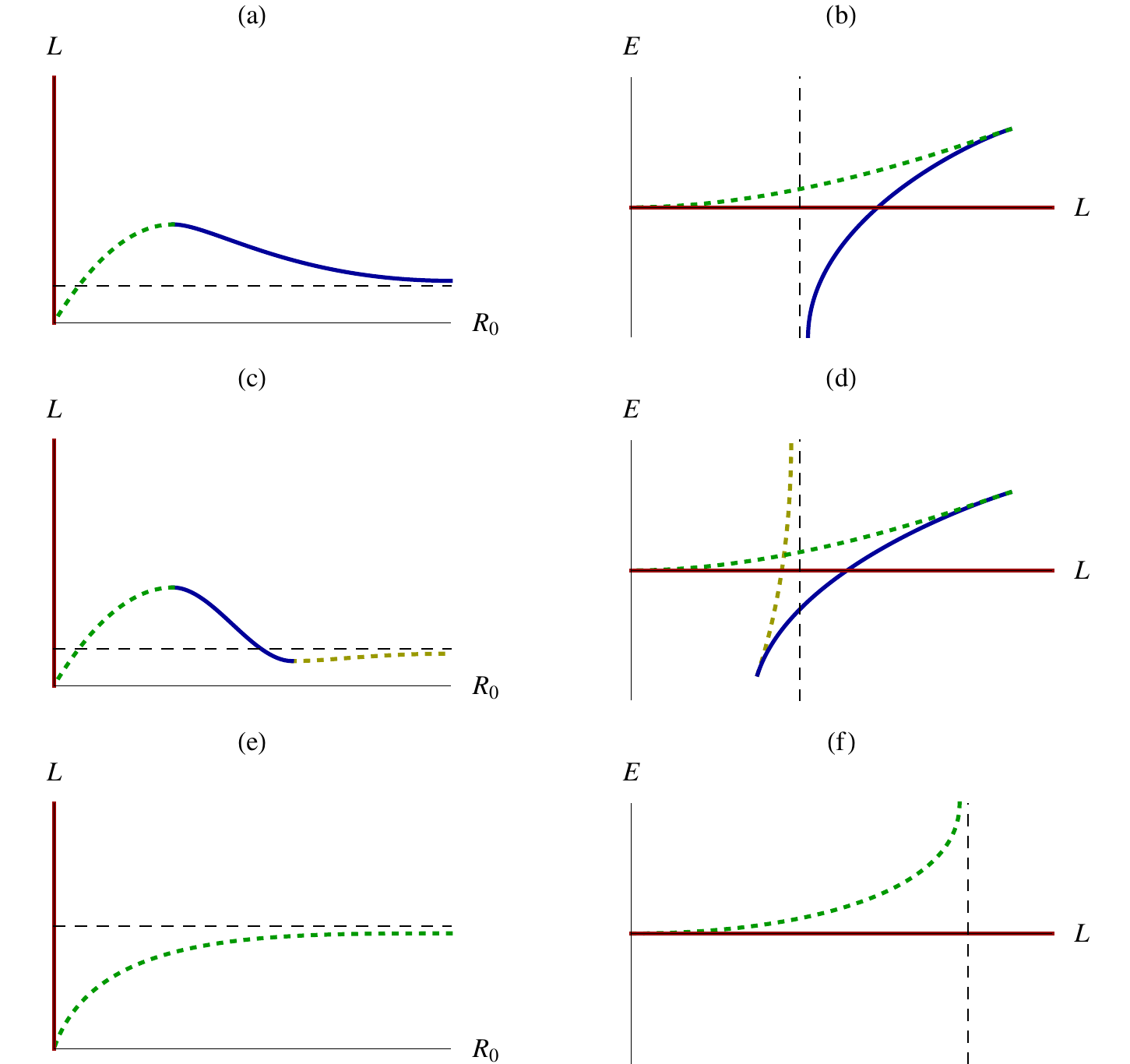}
	\caption{The qualitative behaviour of $L(\Rmin)$ and $E(L)$ in the `soap film' case when $L(\Rinf)\ne0$.}
	\label{fig:linf}
\end{figure}

The generalisation of the discussion of \autoref{sec:energystab} to account for non-zero $\Linf$ is simple. In general there is a region with $L<\Linf$ in which the `free' solution \eqref{eq:zeroSol} is the stable one, as in \autoref{fig:linf}. If there is a minimum, so that $L(R)$ approaches the asymptote from below, then there is an additional (presumably unstable) branch as in \autoref{fig:linf} (c,d).

However, in some cases $L(R)$ is always increasing, as in \autoref{fig:linf} (e,f). In this case the considerations discussed in \autoref{sec:energystab} suggest that the `free' solution is the only stable one for all $L$.

%% file: examples.tex
\section{Application to specific cases}\label{sec:examples}
The results obtained above are not immediately useful when we are interested in a specific background, as opposed to general considerations. This is because, as was noted in \autoref{sec:rescaleToR}, the integral
\begin{align}
	R(\rho) = \int^\rho \d\rho' \frac{g(\rho')}{f(\rho')}
\end{align}
is either difficult or impossible to evaluate analytically in all but the simplest cases. The results of sections \ref{sec:actionEOM}-\ref{sec:largeR} are easily generalised to allow the more general coordinate $\rho$, by the insertion of factors of
\begin{align}
	\der{R}{\rho}=\frac{g}{f}.
\end{align}
In particular,
\begin{align}
	L(\rhmin) &= 2\int_{\rhmin}^\infty\d\rho\ \frac{g(\rho)}{f(\rho)} \frac{ \Teff(\rhmin) }{\sqrt{\Teff(\rho)^2 - \Teff(\rhmin)^2  }},		\label{eq:L(rho)}			\\
	E(\rhmin) &= 2\int_{\rhmin}^\infty \d\rho\ \frac{g(\rho)\Teff(\rho)}{f(\rho)} 
																				 \left[ \frac{\Teff(\rhmin)}
																				 						 {\sqrt{\Teff(\rho)^2 - \Teff(\rhmin)^2 }}
																				 				-1
																				 \right]
							-	2\int_{\rhb}^{\rhmin} \d\rho\ \frac{g(\rho)\Teff(R)}{f(\rho)}.		\label{eq:energyrho}
\end{align}
Similarly, \eqref{eq:LinfR} is more conveniently written
\begin{align}
	\Linf	=	\pi \lim_{\rho\to\infty} \frac{\Teff(\rho)}{\Teff'(\rho)}\frac{g(\rho)}{f(\rho)}.		\label{eq:LinfRho}
\end{align}

\subsection{The Klebanov-Strassler model}
To demonstrate these ideas, we will look at the case of a string in the Klebanov-Strassler background \citep{Klebanov:2000hb}. This is a convenient, relatively simple, example of a theory which results in confining behaviour for large $L$.

If we restrict our attention to 1-dimensional objects, we need only the non-compact part of the metric,
\begin{align}
	ds^2 = h^{-1/2} \d x_{1,3}^2 + \frac16 \epsilon^{4/3} \frac{h^{1/2}}{K^2} \d\rho^2	+ \dsint,		\label{eq:KSmetric}
\end{align}
where
\begin{align}
	h(\rho)		&=	\alpha \frac{2^{2/3}}{4} \int_\rho^\infty \d x\ \frac{x\coth x-1}{\sinh^2 x} ( \sinh 2x - 2x )^{1/3},		\label{eq:KSh}\\
	K(\rho)^3	&=	\frac{\sinh 2\rho -2\rho}{2 \sinh^3 \rho}.		\label{eq:KSK}
\end{align}
The functions appearing in the action \eqref{eq:1dAction} are
\begin{align}
	f(\rho)^2 &= h^{-1},			\nn\\
	g(\rho)^2 &= \frac{\epsilon^{4/3}}{6K^2},
\end{align}
while for a string $\Teff=f$.

For small $\rho$, we have
\begin{align}
	h	=	h_0	-	h_2	\rho^2	+	\cdots,		\qquad\qquad
	g	\sim \frac{1}{K} =	g_0	+	g_2	\rho^2	+	\cdots.
\end{align}
Then the contribution from the lower limit of the integral \eqref{eq:L(rho)} is
\begin{align}
	L(\rhmin)	=	\int_{\rhmin} \frac{g_0h_0}{\sqrt{h_2}} \frac{1}{\sqrt{\rho^2-\rhmin^2+\cdots}},
\end{align}
so that $L$ diverges logarithmically for $\rhmin\to0$. Using \eqref{eq:dEdL}, we find that for small $\rhmin$
\begin{align}
	E'(L) \to \Teff(0) = \text{constant},
\end{align}
as expected for a confining theory.

The results described in \autoref{sec:largeR} have limited application to this case, simply telling us that $L$ does not approach a non-zero constant for large $\rho$. The relevant function \eqref{eq:LinfRho} is for large $\rho$
\begin{align}
 \frac{\Teff(\rho)}{\Teff'(\rho)}\frac{g(\rho)}{f(\rho)} = \frac{g(\rho)}{f'(\rho)} \sim \sqrt{\rho}e^{-\rho/3},
\end{align}
which is not constant.

\begin{figure}[htbp]
	\centering
		\includegraphics{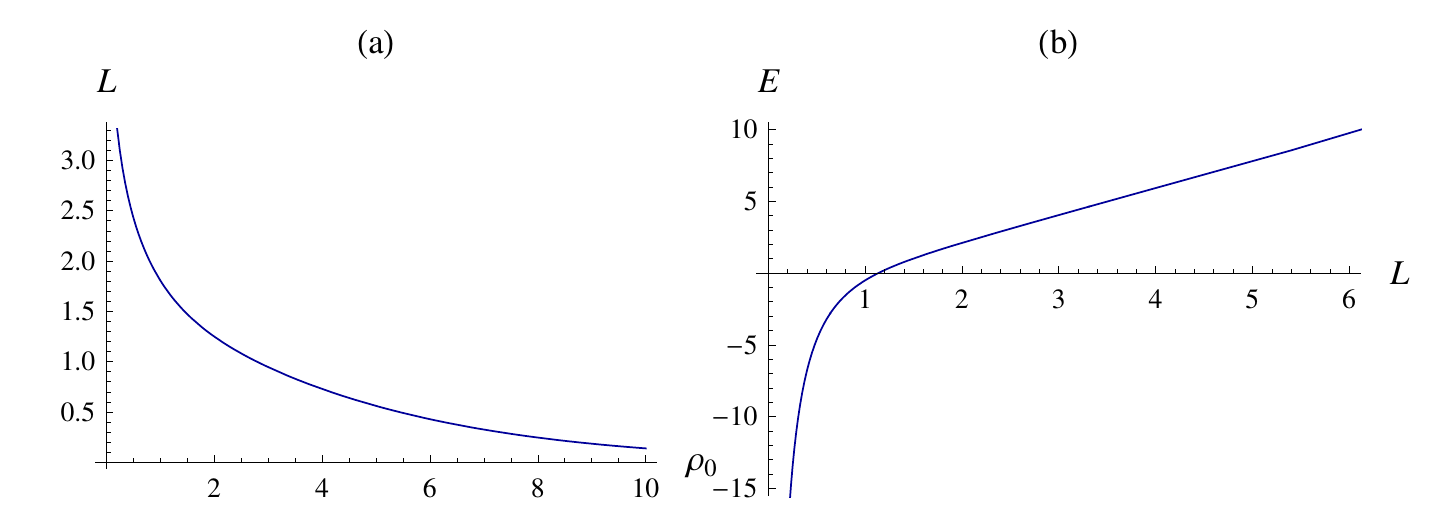}
	\caption{Plots of (a) $L(\rhmin)$ and (b) $E(L)$ for a string in the Klebanov-Strassler background (\ref{eq:KSmetric}-\ref{eq:KSK}), obtained numerically.}
	\label{fig:KSStringEL}
\end{figure}

To obtain the behaviour over the full range of $\rho$ it is necessary to integrate (\ref{eq:L(rho)}-\ref{eq:energyrho}) numerically. The result is shown in \autoref{fig:KSStringEL}. The expected confining behaviour is seen for large $L$, and the form is qualitatively that of \autoref{fig:confining}.

\subsection{$AdS_5\text{-Schwarzschild}\times S^5$}\label{sec:AdSSchwarz}
An example which produces the `soap film' behaviour (\autoref{fig:soapfilm}) is the large mass limit of the $AdS_5\text{-Schwarzschild}$ black hole, expected to describe finite-temperature $\cN=4$ SYM \citep{Witten:1998zw}. This was discussed in \citep{Arias:2009me}.

The metric is
\begin{align}
	ds^2	=	\frac{r^2}{\cR^2} \left[	-\left(	1 - \frac{\mu^4}{r^4}	\right)\d t^2		+	\d\vec{x}^2
														\right]
					+\frac{\cR^2}{r^2} \left(	1 - \frac{\mu^4}{r^4}	\right)^{-1} \d r^2
					+\cR^2 \d\Omega_5^2,	\label{eq:AdSSchwMetric}
\end{align}
so that $g(r)=1$ and
\begin{align}
	f(r)	=	\frac{1}{\cR^2} \sqrt{r^4-\mu^4}.
\end{align}
The horizon is at $\rb = \mu$, and we will consider strings, for which $\Teff=f$. Then $\Teff(r)\to0$ for $r\to\rb$, as is necessary for the`soap film' behaviour.

This background is in fact simple enough that we can obtain the rescaled coordinate $R(r)$ exactly, although not in a form which is particularly useful. We have
\begin{align}
	\der{R}{r} = \frac{g}{f} = \frac{\cR^2}{\sqrt{r^4-\mu^4}},
\end{align}
and so we can define
\begin{align}
	R(r)	\equiv	\cR^2\int_\mu^r \frac{\d r'}{\sqrt{{r'}^4-\mu^4}}
				=	\frac{\sqrt{\pi}\Gamma\left(\frac54\right)}{\Gamma\left(\frac34\right)}
					\frac{\cR^2}{\mu}
					-	\frac{\cR^2}{r} \hyF\left( \frac14 , \frac12; \frac54; \frac{\mu^4}{r^4} \right).
\end{align}
This results in $R(\rb)=0$ and
\begin{align}
	\Rinf \equiv R(r\to\infty) = \frac{\sqrt{\pi}\Gamma\left(\frac54\right)}{\Gamma\left(\frac34\right)}
																\frac{\cR^2}{\mu}.
\end{align}

To find the behaviour for large $r$ we need the function
\begin{align}
	\frac{T(r)}{T'(r)}\frac{g}{f} = \frac{\cR^2}{2} \frac{\sqrt{r^4-\mu^4}}{r^3}	\underset{r\to\infty}{\to} \frac{\cR^2}{2r}.
\end{align}
Referring to \autoref{sec:largeR} we see that this will result in $L(\rmin)\to0$ for $\rmin\to\infty$.

\begin{figure}[htbp]
	\centering
		\includegraphics{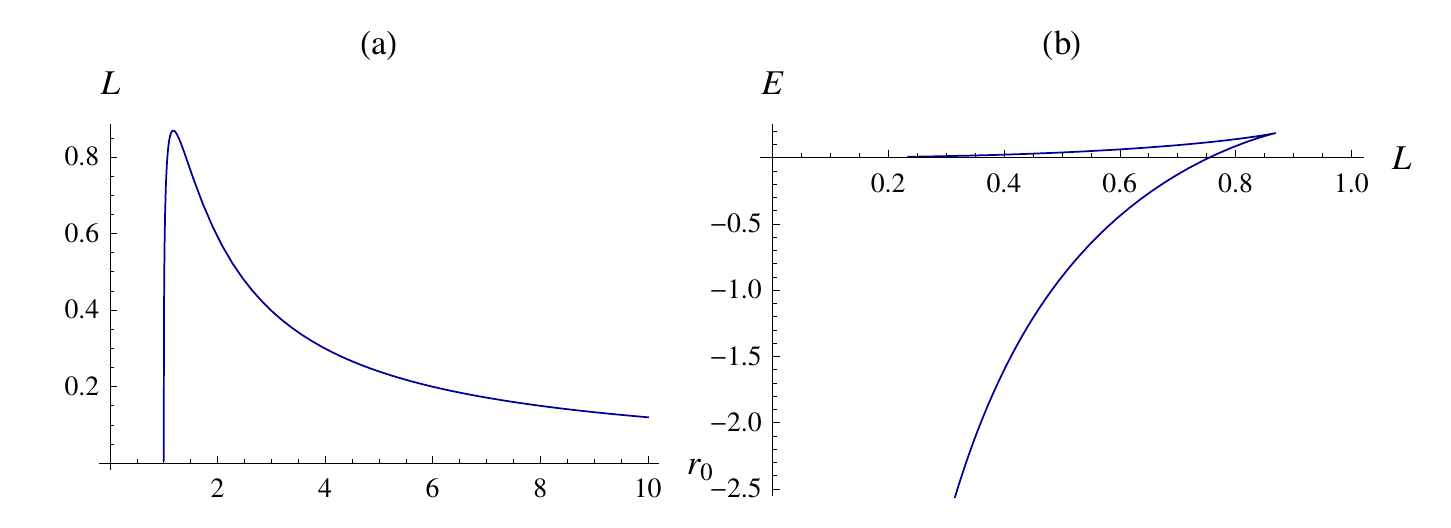}
	\caption{Plots of (a) $L(\rmin)$ and (b) $E(L)$ for a string in the $AdS_5\text{-Schwarzschild}\times S^5$ background \eqref{eq:AdSSchwMetric}, for $\mu = \cR = 1$.}
	\label{fig:AdSSStringEL}
\end{figure}

We can write the separation in the form of \eqref{eq:LintdT}, as
\begin{align}
	L(\rmin)	=	\cR\Teff(\rmin) \int_{\Teff(\rmin)}^\infty\d\Teff \left[\Teff^2 + \frac{\mu^4}{\cR^4}\right]^{-3/4}
																										\frac{1}{\sqrt{\Teff^2-\Teff(\rmin)^2 }}	\label{eq:AdSSchwL}
\end{align}
For small $r$, $\Teff(r)\to0$, and so the lower limit of the integral contributes
\begin{align}
	L(\rmin)	= \cR \Teff(\rmin) \int_{\Teff(\rmin)} \frac{\cR^3}{\mu^3} \frac{1}{\sqrt{\Teff^2 -\Teff(\rmin)^2}}
						\sim \Teff(\rmin) \log \Teff(\rmin) \underset{\rmin\to\rb}{\to} 0.
\end{align}
It is actually possible to evaluate \eqref{eq:AdSSchwL} exactly, resulting in
\begin{align}
	L(\rmin) = \frac{2\sqrt{\pi} \Gamma\left(\frac34\right)}{\Gamma\left(\frac14\right)}
							\frac{ \cR^2 }{ (r^4-\mu^4)^{1/4} }
							\hyF\left(	\frac34, \frac34; \frac54; -\frac{\mu^4}{r^4-\mu^4}
										\right).
\end{align}
Together with the function $E(L)$, obtained numerically, this is shown in \autoref{fig:AdSSStringEL}. The relationship with \autoref{fig:soapfilm} is clear.

%% file: NewSol.tex
\section{The flavoured resolved deformed conifold}\label{sec:NewSol}
Having discussed the necessary techniques and general results, we can turn to the main new material presented in this paper. The analysis of \citep{Maldacena:2009mw} describes a system of solutions related by a chain of dualities, together with a boost in eleven dimensions. More specifically, the `unrotated' solution, obtained by setting the boost parameter $\beta=0$, corresponds D5 branes wrapping the $S^2$ of the resolved conifold \citep{Casero:2006pt}, and is a simpler limit of the solution with general $\beta$ as introduced in \citep{Butti:2004pk}. This has additional D3 brane charges which are not present in the `unrotated' case.

Taking the limit $\beta\to\infty$, we obtain the `rotated' solution, which describes the baryonic branch of the Klebanov-Strassler theory.

The metric is of the form
\begin{align}
		ds^2		&=	h^{-1/2} \d x_{1,3}^2 + e^{2\Phi}h^{1/2} ds_6^2.
\end{align}
The warp factor is related to the dilaton by
\begin{align}
	h = e^{-2\Phi} - e^{-2\Phin}\tanh{\beta},		\label{eq:roth}
\end{align}
where $\Phin$ is the asymptotic value of $\Phi$ for large $\rho$.

These expressions also apply in the flavoured generalisation, described in \citep{Gaillard:2010qg}. However, the $\rho$-dependence of the functions is different. Most significantly, the solutions are now singular in the IR: $\Phi\to-\infty$ for $\rho\to0$.

Turning to the UV, we find that to leading order $\Phi(\rho)$ is unchanged by the addition of flavours. However, the form of \eqref{eq:roth} is such that in the `rotated' case ($\beta=1$) is such that $h(\rho)$ is sensitive to the sub-leading behaviour of $\Phi$. The `rotated' solution therefore has different UV asymptotics in the flavoured case. This is interpreted in \citep{Gaillard:2010qg} as resulting from smeared source D3 branes, uniformly distributed in $\rho$. These result from the action of the `rotation' on the source D5 branes, in the same way as in the unflavoured case the `rotated' solution has bulk D3 branes resulting from the colour D5 branes in the `unrotated' solution.

The aim here is to use the methods discussed in the previous sections to assess the physical significance of these changes with respect to the unflavoured solutions. The changes in the UV asymptotics can be isolated by considering strings which do not descend close to $\rho=0$. The results derived in \autoref{sec:largeR} will therefore apply.

It appears more difficult to isolate the effects of the IR singularity, as the string always probes the large $\rho$ region as well. However, in most cases we will find that the effective tension $\Teff$ vanishes for small $\rho$. As discussed in \autoref{sec:energystab}, this means that only the lower limit of the integral \eqref{eq:L(R)} contributes to $L(\rho)$. The limiting behaviour of $L(\rhmin)$ for small $\rhmin$ is therefore presumably insensitive to changes to the UV.

\subsection{The solutions}
We will now define the solutions of interest more concretely. The metric is of the form
\begin{align}
	ds^2		&=	h^{-1/2} \d x_{1,3}^2 + e^{2\Phi}h^{1/2} ds_6^2,			\nn\\
	ds_6^2	&=	e^{2k} \d\rho^2	+	e^{2q} (\w_1^2 + \w_2^2)
							+	\frac{1}{4} e^{2g} \left[	(\wt_1+a\w_1)^2 + (\wt_2-a\w_2)^2	\right]
							+ \frac{1}{4} e^{2k}	(\wt_3+\w_3)^2,		\label{eq:newSolMetric}
\end{align}	
where
\begin{center}
		\begin{tabular}{r @{\ } c @{\ } l @{\qquad\qquad\qquad} 	r @{\ } c @{\ } l}
		$\w_1$	&	$=$	&	$\d\q$, 	 					&			$\wt_1$	&	$=$	&	$\cos\psi\,\d\qt	+	\sin\psi\,\sin\qt\,\d\vpht$,		\\
		$\w_2$	&	$=$	&	$\sin\q \,\d\vph$,	&			$\wt_2$	&	$=$	&	$-\sin\psi\,\d\qt	+	\cos\psi\,\sin\qt\,\d\vpht$,		\\
		$\w_3$	&	$=$	&	$\cos\q \,\d\vph$,	&			$\wt_3$	&	$=$	&	$\d\psi	+	\cos\qt\, \d\vpht$.		\\
		\end{tabular}
\end{center}
The coefficient functions $\{ \Phi, g, k, q, a\}$ depend only on $\rho$, and as above
\begin{align}
	h = e^{-2\Phi} - e^{-2\Phin}\tanh{\beta}.
\end{align}

The coefficient functions were shown in \citep{Casero:2007jj,HoyosBadajoz:2008fw} to be given by
\begin{align}
	e^{2q} = \frac{1}{4}\frac{P^2-Q^2}{P\cosh\tau - Q},		\qquad\qquad
	e^{2g} = P\cosh\tau -Q,		\qquad\qquad
	a			 = \frac{P\sinh\tau}{P\cosh\tau-Q},
\end{align}
and the BPS equations reduce (after some choices for constants of integration) to 
\begin{align}
	\sinh\tau &=	\frac{1}{\sinh2\rho},										\nn\\
	Q					&=	\frac{2\Nc-\Nf}{2} (2\rho\cosh\tau -1),	\nn\\
	e^{4(\Phi-\Phio)}
						&=	\frac{4}{(P^2-Q^2)e^{2k}\sinh^2\tau},		\nn\\
	e^{2k}		&=	\frac{1}{2}(P'+\Nf),													\label{eq:newSolFunc}
\end{align}
together with
\begin{align}
	P'' + (P'+\Nf) \left(		\frac{P'+Q'+2\Nf}{P-Q}
												+	\frac{P'+Q'+2\Nf}{P+Q}
												-	4\cosh\tau
									\right)													=0.		\label{eq:mastereq}
\end{align}
The solution discussed in \citep{Gaillard:2010qg} is given by the asymptotic behaviour of $P(\rho)$,
\begin{align}
	P = \begin{cases}
					h_1\rho + \frac{4\Nf}{3} \left[	-\rho \log\rho -\frac{1}{12}\rho\log(-\log\rho)
																					+	\cO\left( \frac{\rho\log(-\log\rho)}{\log\rho} \right)
																	\right]	+\cO(\rho^3\log\rho),					& \rho\to0						\\ \\
					ce^{4\rho/3}	+ \frac{9\Nf}{8}
												+ \frac{1}{c} \left[	(2\Nc-\Nf)^2\left(\rho^2-\rho+\frac{13}{16}\right)
																							- \frac{81\Nf^2}{64}
																			\right]	e^{-4\rho/3}
												+ \cO(\rho e^{-8\rho/3} ),												& \rho\to\infty,
			\end{cases}	\label{eq:Pasymptotics}
\end{align}
where the two arbitrary constants $h_1$ and $c$ are related in a non-trivial way.

The full solution to \eqref{eq:mastereq} can then be found numerically, interpolating between the two regimes in \eqref{eq:Pasymptotics}.

\begin{figure}[htbp]
	\centering
		\includegraphics{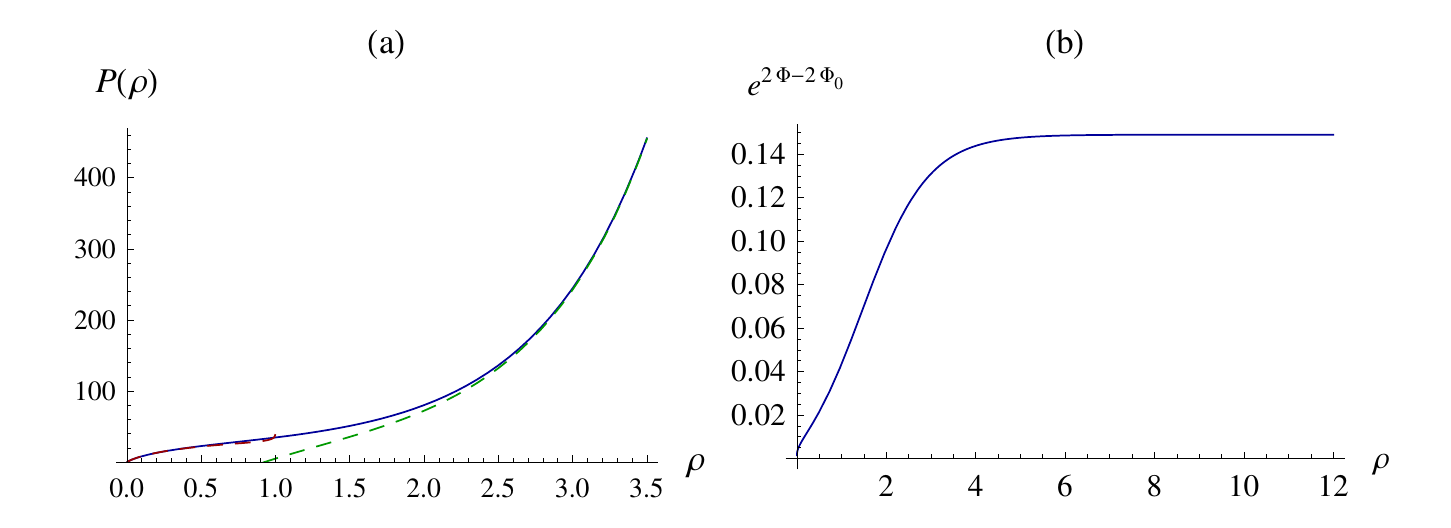}
	\caption{The generic behaviour of (a) $P(\rho)$ and (b) $\Phi(\rho)$, obtained numerically. The red and green dashed curves show the first few terms of the expansions \eqref{eq:Pasymptotics} for small and large $\rho$ respectively.}
	\label{fig:rotatedPPhi}
\end{figure}

%% file: rotated.tex
\subsection{The `rotated' case}\label{sec:rot}
The functions appearing in the action \eqref{eq:1dAction} are
\begin{align}
	f(\rho)^2 &= h^{-1}	= \frac{1}{ e^{-2\Phi} - e^{-2\Phin}\tanh{\beta} },			\nn\\
	g(\rho)^2 &= e^{2\Phi+2k}.
\end{align}
To obtain $f(\rho\to\infty)\to\infty$, as required by the boundary conditions for a string (\autoref{sec:UVBCs}) we therefore require the limit $\beta\to\infty$. This is the `rotated' solution; the field theory limit discussed in \citep{Gaillard:2010qg}.

\subsubsection{Fundamental string}
In the case of a fundamental string we simply have $\Teff=f/2\pi\alpha'$. We first consider strings which descend deep into the space. For small $\rho$, the functions we need have the asymptotic behaviour \citep[Appendix B]{Gaillard:2010qg}
\begin{align}
	e^{4(\Phi-\Phio)}	&=	\frac{27}{2\Nf^3}(-\log\rho)^{-3}
												\left[ 1 + \cO\left( \frac{\log(-\log\rho)}{-\log\rho} \right)\right],	\label{eq:dilatonIR}		\\
	e^{2k}						&=	\frac{2\Nf}{3} (-\log\rho)
												\left[ 1 + \cO\left( \frac{\log(-\log\rho)}{-\log\rho} \right)\right].		\label{eq:kIR}
\end{align}
Writing $2\pi\alpha'=1$ this results in
\begin{align}
	\Teff(\rho) = f(\rho) = \left( \frac{27}{2\Nf^3} \right)^{1/4} e^{\Phio}(-\log\rho)^{-3/4}
								\left[ 1 + \cO\left( \frac{\log(-\log\rho)}{-\log\rho} \right)\right].					
\end{align}
In particular, $\Teff(0)=0$, so we generically expect qualitatively `soap film' behavior as in \autoref{fig:soapfilm}, or one of the modifications discussed subsequently. The `free' solution \eqref{eq:zeroSol}, with $\rhmin=0$, exists for all $L$. In the case of the the smooth solution \eqref{eq:smoothSol} the separation is given, as in \autoref{sec:largeR}, by
\begin{align}
		L(\rhmin)	=	2	\int_1^\infty	\d t\; \frac{t}{t'(R)}
																			 \frac{1}{t\sqrt{	t^2 - 1		}}.
\end{align}
Generalising this to allow us to continue working with $\rho$ rather than $R$, we can write
\begin{align}
		L(\rhmin)	=	2	\int_1^{t(\rho_*)}	\d t\; \frac{t}{t'(\rho)}\frac{g(\rho)}{f(\rho)}
																				 \frac{1}{t\sqrt{	t^2 - 1		}}
									+	2	\int_{t(\rho_*)}^\infty	\d t\; \frac{t}{t'(\rho)}\frac{g(\rho)}{f(\rho)}
																										 \frac{1}{t\sqrt{	t^2 - 1		}},
\end{align}
where we have split the integral at an arbitrary point $\rho=\rho_*$
As $\Teff(0)=0$, the limit $\rhmin\to0$ results in
\begin{align}
	t(\rho_*) \equiv \frac{\Teff(\rho_*)}{\Teff(\rhmin)} \to \infty.
\end{align}
The separation is then
\begin{align}
	L(\rhmin\to0) = 2	\int_1^\infty	\d t\; \frac{t}{t'(\rho)}\frac{g(\rho)}{f(\rho)}
																				 \frac{1}{t\sqrt{	t^2 - 1		}}.								\label{eq:Lrhmin}
\end{align}
This integral covers only the range $0<\rho<\rho_*$. As $\rho_*$ was arbitrary, we can take the limit $\rho_*\to0$ and evaluate \eqref{eq:Lrhmin} exactly using the small-$\rho$ asymptotic expression for $tg/t'f$. Then, using (\ref{eq:dilatonIR}-\ref{eq:kIR}),
\begin{align}
	\frac{t(\rho)}{t'(\rho)}\frac{g(\rho)}{f(\rho)} = \frac{g(\rho)}{f'(\rho)}
											=	\frac{4}{3}\sqrt{\frac{2\Nf}{3}}\rho(-\log\rho)^{3/2}
												\left[ 1 + \cO\left(\frac{\log(-\log\rho)}{-\log\rho} \right)\right].
\end{align}
We also have
\begin{align}
	t	=	\frac{f(\rho)}{f(\rhmin)}	=	\left( \frac{\log\rhmin}{\log\rho} \right)^{3/4}	+	\cdots,
\end{align}
so
\begin{align}
	\frac{tg}{t'f}	=	\frac{4}{3} \sqrt{\frac{2\Nf}{3}}\rhmin(-\log\rhmin)^{3/2} e^{t^{-4/3}}t^{-2} + \cdots,
\end{align}
and
\begin{align}
	L(0)	=	\frac{4}{3}\sqrt{\frac{2\Nf}{3}}	\left[ \lim_{\rhmin\to0}\rhmin(-\log\rhmin)^{3/2}
					\int_1^\infty	\d t\ \frac{e^{t^{-4/3}}}{t^3\sqrt{t^2-1}} \right].
\end{align}
The integral is finite, so $L(0)=0$, as in \autoref{fig:soapfilm} and \autoref{fig:linf}.

We now turn to the behaviour of strings with large $\rhmin$. For large $\rho$, the metric functions are
\begin{align}
	e^{4(\Phi-\Phin)}	&=	1	-	\frac{3\Nf}{c} e^{-4\rho/3}
													+	\frac{3}{16c^2}\left[	(2\Nc-\Nf)^2 (1-8\rho) + 297\Nf^2 \right] e^{-8\rho/3}
													+ \cO\left(e^{-4\rho}\right),\label{eq:rotatedDilatonUV}\\
	e^{2k}						&=	\frac{2c}{3}e^{4\rho/3} \left[ 1	+	\frac{3\Nf}{4c}e^{-4\rho/3}
																													+	\cO\left(\rho^2 e^{-8\rho/3}\right)
																								\right],
\end{align}
where
\begin{align}
	e^{2\Phin} = \sqrt{\frac32}\frac{e^{2\Phio}}{c^{3/2}}.
\end{align}
This results in
\begin{align}
	f	&=\sqrt{\frac{2c}{3\Nf}} e^{\Phin} e^{2\rho/3}
			\left\{		1		+		\frac{1}{32c\Nf} \left[	(2\Nc-\Nf)^2(1-8\rho) + 216\Nf^2	\right]	e^{-4\rho/3}
										+		\cO\left(e^{-8\rho/3}\right)
			\right\},			\label{eq:rotUVf}\\
	g	&=\sqrt{\frac{2c}{3}}e^{\Phin} e^{2\rho/3} \left[1	-	\frac{3\Nf}{8c}e^{-4\rho/3}
																												+	\cO\left(\rho^2 e^{-8\rho/3} \right)
																							\right].
\end{align}
We again need the function $tg/t'f$, which for large $\rho$ is
\begin{align}
	\frac{tg}{t'f} =	\frac{3}{2}\sqrt{\Nf}	-	\frac{3}{64c\sqrt{\Nf}}\left[ (2\Nc-\Nf)^2(8\rho+11) - 249\Nf^2 \right]
																						e^{-4\rho/3}
																					+	\cO\left( \rho^2 e^{-8\rho/3} \right).	\label{eq:rotatedStringLinf}
\end{align}
Using \eqref{eq:LinfRho}, for large $\rho$ the separation therefore approaches
\begin{align}
	\Linf = \frac{3\pi}{2}\sqrt{\Nf}.
\end{align}
If we include the next term of the expansion \eqref{eq:rotatedStringLinf} in the integral \eqref{eq:Lintdt}, we get
\begin{align}
	L(\rhmin) = \Linf - 2 \int_1^\infty \d t\ \frac{\epsilon(\rho(t,\rhmin))}{t\sqrt{t^2-1}}	+	\cdots,
\end{align}
where
\begin{align}
	\epsilon(\rho) = \frac{3(2\Nc-\Nf)^2}{8c\sqrt{\Nf}}  \rho e^{-4\rho/3} >0.
\end{align}
This means that $L$ approaches $\Linf$ from below. Together with the fact that the smooth solution has $L(0)=0$, the result is behaviour which is qualitatively that of \autoref{fig:linf} (c-f).

For the behaviour of $E$, we need the function $\ex'(\Teff)$, defined in \eqref{eq:defexprime}. In this case, for large $\rho$ we simply have
\begin{align}
	\ex'(\Teff) &= - \epsilon(\rho(\Teff)) + \cO(e^{-4\rho/3})		\nn\\
							&= - \frac{3(2\Nc-\Nf)^2}{8\Nf}\sqrt{\frac{3}{2c}} \frac{\log\Teff}{\Teff^2}
										+	\cO\left( \frac{1}{\Teff^2} \right).
\end{align}
This satisfies the condition \eqref{eq:condexprime}, so the energy approaches a constant for large $\rho$. More precisely, \eqref{eq:E(ex)} results in
\begin{align}
	E(\rhmin) = \Einf -	\frac{3(2\Nc-\Nf)^2}{4\Nf}\sqrt{\frac{3}{2c}}
																(2\pi-3)
																\frac{\log(\Teff(\rhmin))}{\Teff(\rhmin)}
															+	\cO\left(	\frac{1}{\Teff(\rhmin)} \right),
\end{align}
where
\begin{align}
	\Einf \equiv 2\ex(0) = 2 \int_0^\infty \d\Teff \left[ \frac{\Linf}{\pi} - \frac{\Teff}{\Teff'(R)} \right]
\end{align}
is a constant.

\begin{figure}[htbp]
	\centering
		\includegraphics{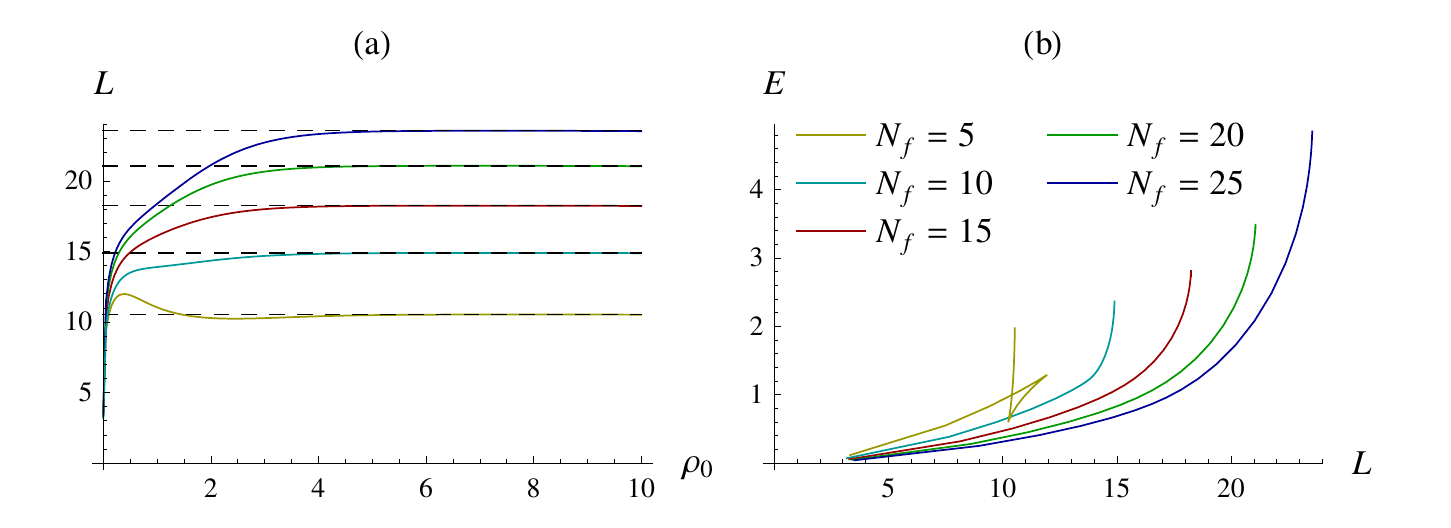}
	\caption{The results of the numerical calculation of (a) $L(\rhmin)$ and (b) $E(L)$ for a string in the `rotated' background, for various values of $\Nf$. Here $\Nc=10$ and $h_1=25$, resulting in values of $c$ in the range $2.8\lesssim c\lesssim 5.5$, depending on $\Nf$.}
	\label{fig:rotatedStringEL}
\end{figure}

The numerical calculation (\autoref{fig:rotatedStringEL}) confirms this, and reveals that a local maximum and minimum occur in $L(\rho)$ for small $\Nf$, while for large $\Nf$ we find that $L(\rho)$ is always increasing, so that $E(L)$ is smooth. In all the cases calculated $E>0$ for all $L$, and so the stable configuration is presumably the `free' solution \eqref{eq:zeroSol}.

Note that because a finite upper limit (in this case $\rhoM=20$) is needed for the integrals, the numerical calculations cannot be trusted for large $\rho$. In particular, the numerical integration of \eqref{eq:L(rho)} will always yield $L(\rhoM)=0$. The plots in \autoref{fig:rotatedStringEL} have been terminated before $L$ decreases significantly away from $\Linf$.

\begin{figure}[htbp]
	\centering
		\includegraphics{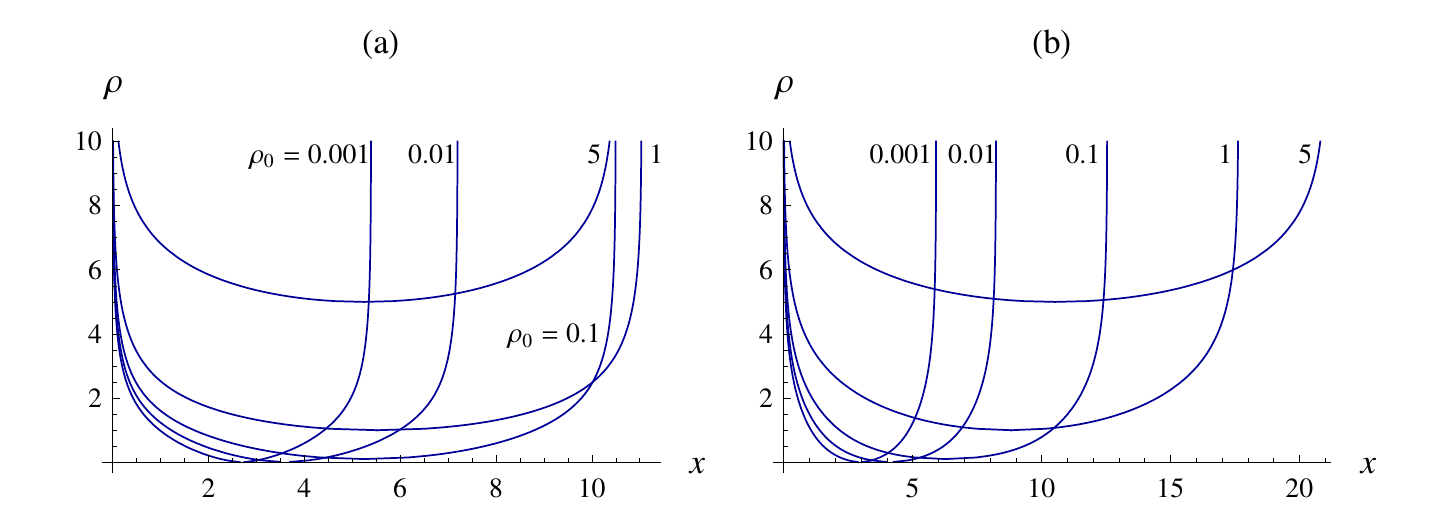}
	\caption{The shapes of strings with varying $\rhmin$. Here the parameters are as in \autoref{fig:rotatedStringEL}, with (a) $\Nf=5$ and (b) $\Nf=20$.}
	\label{fig:rotatedStringProfile}
\end{figure}

By evaluating \eqref{eq:smoothSol} numerically, we can determine the shape of the string. This is shown in \autoref{fig:rotatedStringProfile}. Of the strings shown all are unstable, except that with $\Nf=5$, $\rhmin=1$, which falls within the region between the cusps in $E(L)$ (\autoref{fig:rotatedStringEL}) and so is metastable.

\subsubsection{D3 brane}\label{sec:rotD3}
The 't Hooft loop can be obtained by replacing the string in the above discussion with a D3 brane \citep{HoyosBadajoz:2008fw}, with
\begin{align}
	\q = \qt,		\qquad\qquad	\vph = 2\pi - \vpht,		\qquad\qquad \psi=\pi.
\end{align}
As described in \autoref{sec:actionEOM}, we therefore need to calculate
\begin{align}
 \Teff(\rho) = \TDt e^{-\Phi} f \int \d^{2}X\; \sqrt{\det G_{ab}^{(2)}},
\end{align}
where
\begin{align}
			G_{ab}^{(2)}		&=	g_{ij}\pder{{\q^i}}{{X^a}}\pder{{\q^j}}{{X^b}}
\end{align}
is the internal part of the induced metric on the D3 brane. If we choose the parametrisation $\{X^2=\q,X^3=\vph\}$, we obtain from \eqref{eq:newSolMetric}
\begin{align}
	G_{ab}^{(2)}\d X^a \d X^b	=	h^{1/2} e^{2\Phi} \left[ e^{2q} + \frac14 e^{2g}(a-1)^2 \right]
																								\left(	\d\q^2	+	\sin^2\q\; \d\vph^2	\right).
\end{align}
Setting $\TDt=1$ for convenience, this results in
\begin{align}
	\Teff	=	4\pi e^{\Phi} \left[ e^{2q} + \frac14 e^{2g}(a-1)^2 \right].		\label{eq:D3Tension}
\end{align}
Again using the asymptotic expansions in \citep{Gaillard:2010qg}, this is
\begin{align}
	\Teff = \begin{cases}
\displaystyle				8\pi e^{\Phio} \rho^2 \left(	-\frac{2\Nf}{3} \log\rho	\right)^{1/4}
																					\left[	1	+	\cO\left(	\frac{\log(-\log\rho)}{\log\rho}	\right)
																					\right],					& \rho\to0,						\\ \\
\displaystyle		2\pi c e^{\Phin} e^{4\rho/3}	\left\{		1	+	\frac{1}{4c}
																														\left[	(2\Nc-\Nf)\rho	+	\frac{3\Nf}{4}	\right]
																														e^{-4\rho/3}
																													+	\cO(e^{-2\rho})			
																							\right\},												& \rho\to\infty.
			\end{cases}	\label{eq:TeffD3asymptotics}
\end{align}
As $\Teff(0)=0$ we generically expect the smooth solution to have $L(0)=0$, as in the case of the string, but to be sure it would again be necessary to evaluate the integral \eqref{eq:Lrhmin}.

To obtain the large $\rhmin$ behaviour, we again need the $\rho$-dependence of $\Teff g/\Teff'f$, and in this case we get an asymptotically constant separation
\begin{align}
	\Linf	=	\frac{3\pi}{4}\sqrt{\Nf},
\end{align}
which is equal to half that obtained for the string.

\begin{figure}[htbp]
	\centering
		\includegraphics{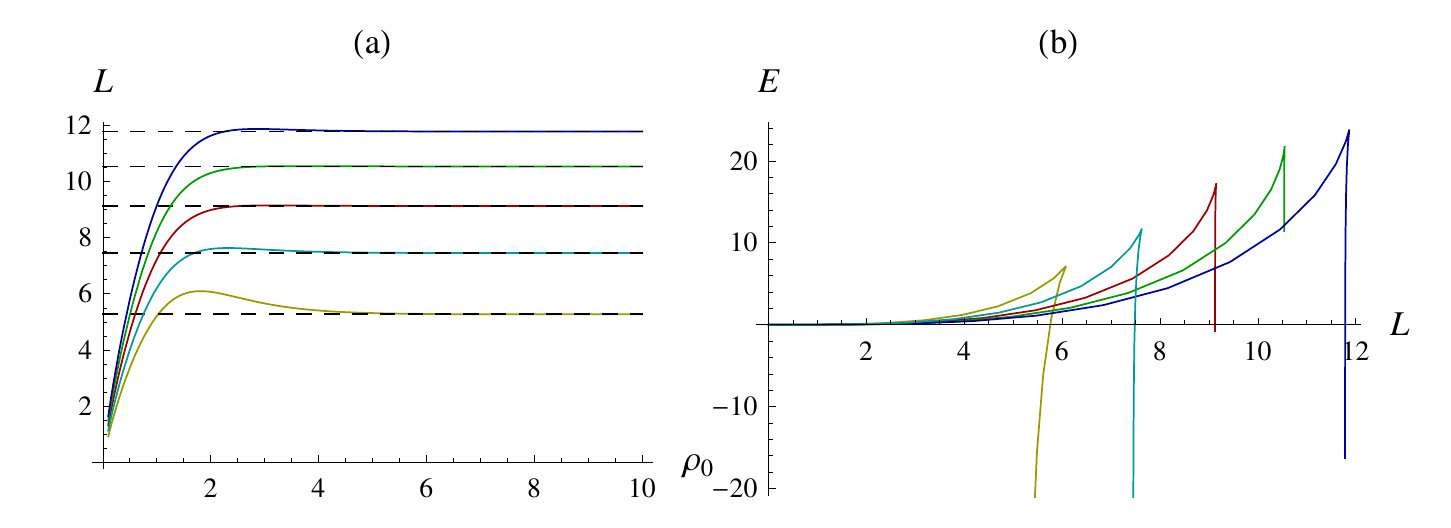}
	\caption{The results of the numerical calculation of (a) $L(\rhmin)$ and (b) $E(L)$ for a D3 brane in the `rotated' background. The parameters have the same values as in \autoref{fig:rotatedStringEL}. Note that the endpoints of the curves were determined by the limitations of the numerical calculation.}
	\label{fig:rotatedD3EL}
\end{figure}

The results of the numerical calculations are shown in \autoref{fig:rotatedD3EL}. The behaviour is qualitatively similar to that of the string (\autoref{fig:rotatedStringEL}), with the exception of the fact that in this case $L$ approaches $\Linf$ from above. Although this appears to be an insignificant difference when viewed in terms of $L(\rhmin)$, the effect is that $E$ decreases for large $\rhmin$. This means that there is a (small) range of $L$ for which the smooth configuration is stable.

\subsubsection{D1 brane}
For a D1 brane the effective tension is
\begin{align}
	\Teff = e^{-\Phi}f,
\end{align}
where we set $\TDo=1$. By \eqref{eq:rotatedDilatonUV}, we see that for large $\rho$ the tension approaches a constant multiple of that of the fundamental string,
\begin{align}
	\Teff^{\text{D1}} \to e^{-\Phin} \Teff^{\text{string}}.
\end{align}
The asymptotic value of $L$ depends on the ratio $\Teff/\Teff'$ and is therefore the same as in the case of the string,
\begin{align}
	\Linf = \frac{3\pi}{2}\sqrt{\Nf}.
\end{align}

For $\rho\to0$, we can write
\begin{align}
	f	=	e^\Phi \left[		1	+	\frac12 e^{2(\Phi-\Phin)}	+	\cO\left( e^{4(\Phi-\Phin)} \right)
							\right],
\end{align}
which results in
\begin{align}
	\Teff = 1 + \frac32 \left( - \frac{\Nf}{c} \log\rho	\right)^{-3/2}
							+	\cO\left(	\frac{\log(-\log\rho)}{(-\log\rho)^{5/2}}	\right).
\end{align}
As $\Teff(0)\ne0$, the integrand in \eqref{eq:L(rho)} is non-zero for $\rho\ne0$, meaning that $L(0)\ne0$. The contribution from the lower limit is of the form
\begin{align}
	\int_0 \d\rho\; \left( -\log\rho \right)^{-3/2},
\end{align}
which is finite. The separation is therefore finite and non-zero for $\rhmin=0$, and the `free' solution \eqref{eq:zeroSol} does not exist. It is unfortunately not possible to determine $L(0)$ analytically, because it would be necessary to evaluate the integral \eqref{eq:L(rho)} over the whole range $0\le\rho<\infty$. 

\begin{figure}[htbp]
	\centering
		\includegraphics{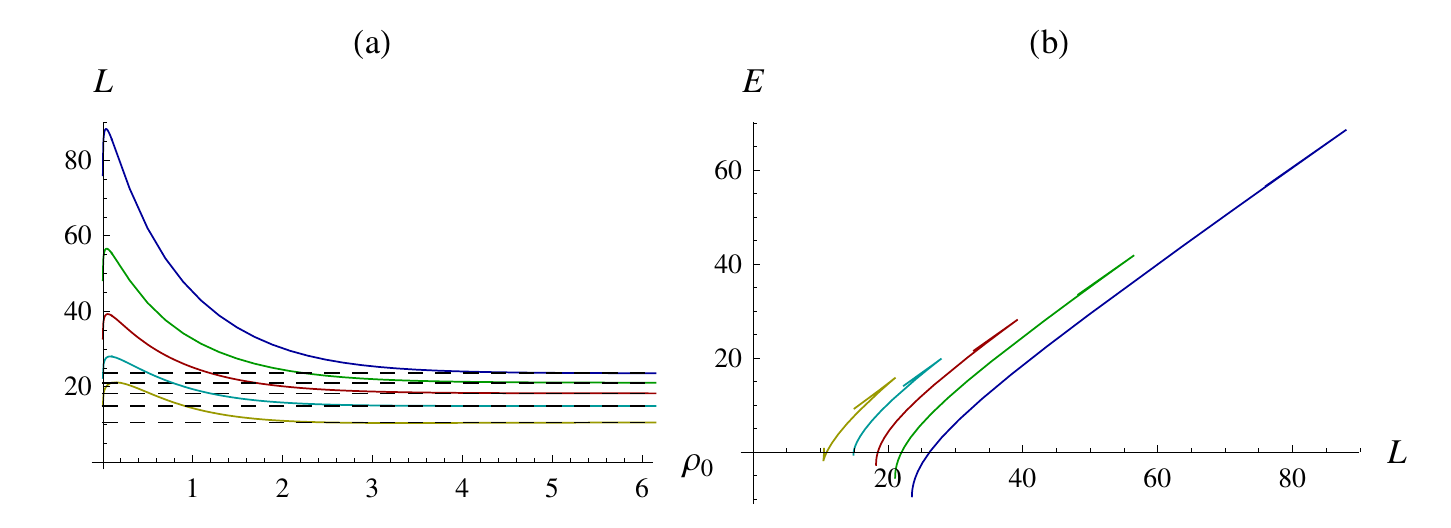}
	\caption{The results of the numerical calculation of (a) $L(\rhmin)$ and (b) $E(L)$ for a D1 brane in the `rotated' background. The parameters have the same values as in \autoref{fig:rotatedStringEL}.}
	\label{fig:rotatedD1EL}
\end{figure}

The results of the numerical calculations are shown in \autoref{fig:rotatedD1EL}. The behaviour for large $\rhmin$ is confining, as expected when $\Teff\ne0$.

%% file: unrotated.tex
\subsection{The `unrotated' case}
In the case with $\beta\to0$, the metric is simply
\begin{align}
	ds^2	=	e^\Phi \left(	\d x_{1,3}^2 + ds_6^2	\right),
\end{align}
and	$f = e^\Phi$. However, generically $\Phi\to\text{constant}$ for large $\rho$, and so the boundary condition \eqref{eq:UVBCs} is not satisfied without $\beta\to\infty$. This problem can be overcome by taking the limit $c\to0$, which can be considered the flavoured generalisation of the solutions discussed in \citep{Maldacena:2000yy}. This case was discussed in \citep{HoyosBadajoz:2008fw}, from which we obtain the asymptotic expansion for large $\rho$,
\begin{align}
	P	=	\abs{2\Nc-\Nf}\rho	\left[	1	+	\frac{\Nf}{2\abs{2\Nc-\Nf}}	\frac{1}{\rho}
																		+	\cO\left(\frac{1}{\rho^2}\right)
													\right].	\label{eq:PasymptoticsLinear}
\end{align}
For small $\rho$ the solution is unchanged from \eqref{eq:Pasymptotics}. Using \eqref{eq:newSolFunc} this results in
\begin{align}
	f	=	e^\Phi	&=	A e^{\Phio}	\rho^{-1/4}e^\rho	\left[	1	+	\cO(\rho^{-1})	\right],	\nn\\
g	=	e^{\Phi+k}&=	\frac{A}{\sqrt{2}} \sqrt{\abs{2\Nc-\Nf}+\Nf}\ \rho^{-1/4}e^\rho	\left[	1	+	\cO(\rho^{-1})	\right],
\label{eq:unrotfg}
\end{align}
where
\begin{align}
	\frac{1}{A^4}	=	\frac{1}{2}(\abs{2\Nc-\Nf}+\Nf)^2\abs{2\Nc-\Nf}.
\end{align}
In this limit we still have $f\to\infty$ for $\rho\to\infty$, so \eqref{eq:UVBCs} is satisfied.
\begin{figure}[htbp]
	\centering
		\includegraphics{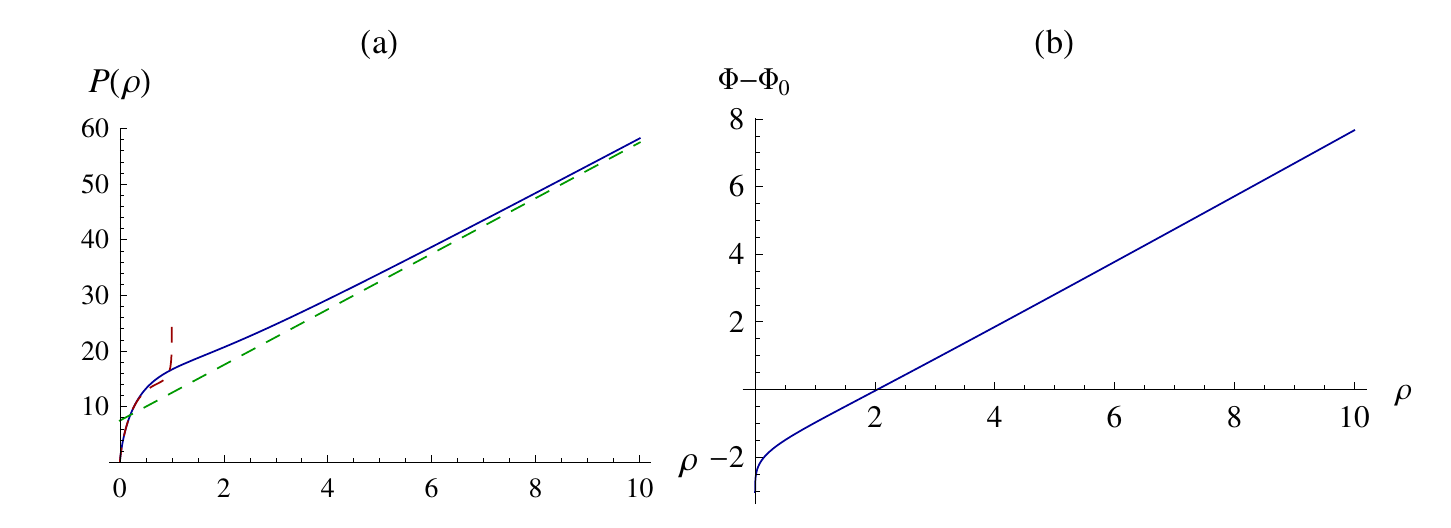}
	\caption{The behaviour of (a) $P(\rho)$ and (b) $\Phi(\rho)$ with $c\to0$, showing the linear behaviour (\ref{eq:PasymptoticsLinear}-\ref{eq:unrotfg}) for large $\rho$.}
	\label{fig:unrotatedPPhi}
\end{figure}

For the purposes of our discussion, the only difference between the `rotated' and `unrotated' backgrounds is
\begin{align}
	f_\text{unrotated}^{-2}	=	e^{-2\Phi},		\qquad\qquad
	f_\text{rotated}^{-2}		=	e^{-2\Phi} - e^{-2\Phin}.
\end{align}
For small $\rho$ we have $\Phi\to-\infty$, and so $f$ is unchanged by the `rotation'. The discussion of the previous section resulting in $L(0)=0$ for the string and D3 brane therefore also applies in the `unrotated' case.

\subsubsection{Fundamental string}
Using \eqref{eq:unrotfg}, and the fact that for a fundamental string $\Teff=f$, we find that for large $\rho$,
\begin{align}
	\frac{tg}{t'f} = \frac{g(\rho)}{f'(\rho)}
								 = \frac{1}{\sqrt{2}}\sqrt{\abs{2\Nc-\Nf}+\Nf} \left[		1	+	\cO(\rho^{-1})	\right],
\end{align}
so that the separation is asymptotically
\begin{align}
	\Linf = \frac{\pi}{\sqrt{2}}\sqrt{\abs{2\Nc-\Nf}+\Nf}.
\end{align}

\begin{figure}[htbp]
	\centering
		\includegraphics{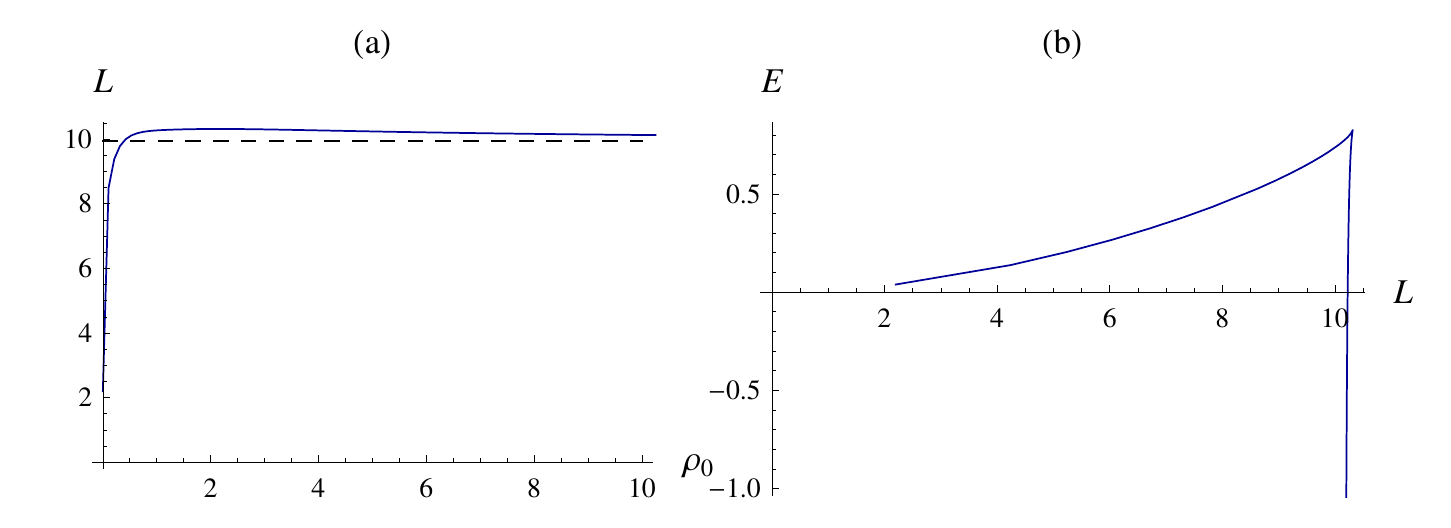}
	\caption{ The results of the numerical calculation of (a) $L(\rhmin)$ and (b) $E(L)$ for a string in the `unrotated' background, with $\Nc=10$ and $\Nf=15$.  To obtain the linear behaviour \eqref{eq:PasymptoticsLinear}, $h_1\approx11.198$, so as to give $c=0$. }
	\label{fig:unrotatedStringEL}
\end{figure}

The numerical calculation again results in the expected modified `soap film' behaviour (as in \autoref{fig:linf}), as shown in \autoref{fig:unrotatedStringEL}.

\subsubsection{D3 brane}
The internal space is unaffected by the `rotation', so the D3 brane effective tension is still given by \eqref{eq:D3Tension}. However, we now have the limit $c\to0$, so the large-$\rho$ asymptotic expression \eqref{eq:TeffD3asymptotics} is no longer valid. Instead, using the asymptotic solutions in \citep{HoyosBadajoz:2008fw}, 
\begin{align}
	\Teff = 2\pi A e^{\Phio} \abs{2\Nc-\Nf} \rho^{3/4} e^\rho \left[ 1	+	\cO\left(\frac{1}{\rho}\right) \right].
\end{align}
Together with \eqref{eq:unrotfg}, this results in
\begin{align}
	\Linf	=	\frac{\pi}{\sqrt{2}}\sqrt{\abs{2\Nc-\Nf}+\Nf},
\end{align}
as for the fundamental string.
\begin{figure}[htbp]
	\centering
		\includegraphics{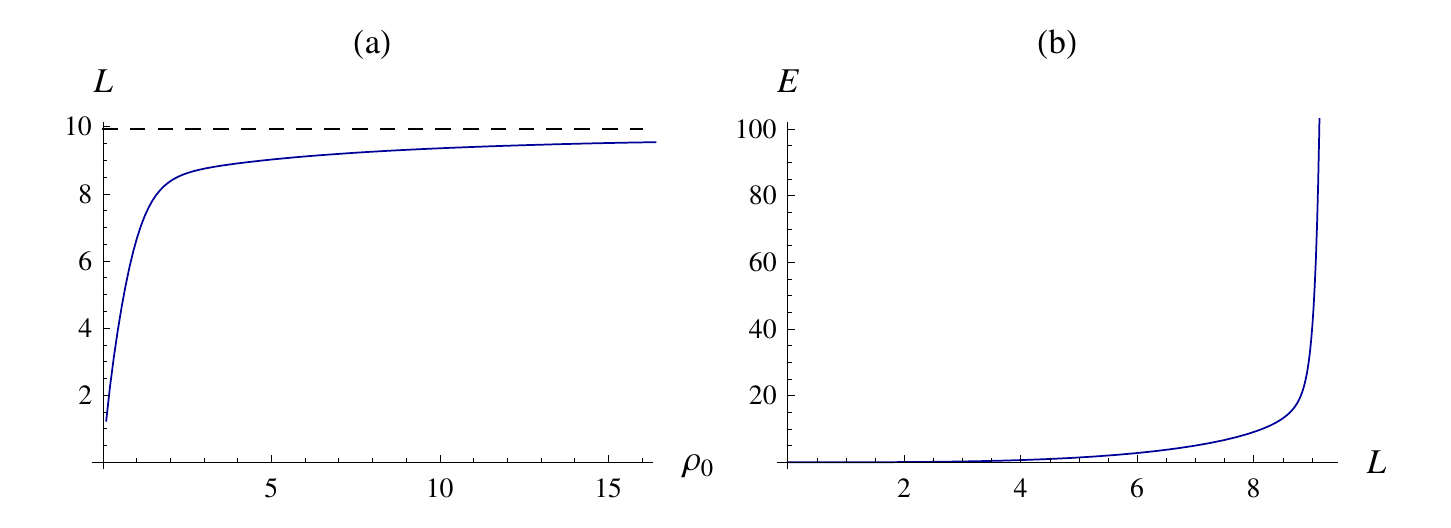}
	\caption{   The results of the numerical calculation of (a) $L(\rhmin)$ and (b) $E(L)$ for a D3 brane in the `unrotated' background. The parameters have the same values as in \autoref{fig:unrotatedStringEL}.   }
	\label{fig:unrotatedD3EL}
\end{figure}

\subsubsection{D1 brane}
In the `unrotated' background a D1 brane has constant tension
\begin{align}
	\Teff = \TDo e^{-\Phi}f = \TDo.
\end{align}
The minimum-energy solution to \eqref{eq:EoM} is then simply a string which does not descend into the bulk, resulting in confining behaviour $E=\TDo L$.

%% file: discussion1.tex
\subsection{Discussion}\label{sec:discussion}

Qualitatively, the clearest result of the preceding analysis is the unusual behaviour seen in the flavoured solutions for small $\rhmin$. For both strings and D3 branes we find  the `soap film' behaviour shown in \autoref{fig:soapfilm}, the primary result of which is that at large separations the only solution to the equation of motion \eqref{eq:EoM} is the `free' solution \eqref{eq:zeroSol}. This not particularly unusual; it is also seen, for example, in the case of the $AdS_5\text{-Schwarzschild}\times S^5$ background discussed in \autoref{sec:AdSSchwarz}.

In \autoref{sec:energystab} we related this behaviour to the fact that the effective tension vanishes for $\Rmin=\Rb$. It seems reasonable to suppose that this change relative to the unflavoured case is related to the introduction of the IR singularity. For example, in the case of the fundamental string we have $\Teff(\rhmin)\sim e^\Phi(\rhmin)$ for small $\rhmin$, and we see that this effect follows immediately from the fact that the dilaton diverges, $\Phi(\rho)\to-\infty$, for small $\rho$.

The modified behaviour for large $\rhmin$ manifests itself in the fact that the separation goes to a non-zero constant for large $\rhmin\to\infty$. This is also simplest when considered in the context of the fundamental string, so that $\Teff\propto f$. In that case the condition \eqref{eq:condForConst} for constant separation in the UV becomes
\begin{align}
	\frac{g(r)}{f'(r)}=\text{constant} 		\qquad\qquad \text{for $r\to\rinf$},
\end{align}
for a generic radial coordinate $r$. In terms of the specific radial coordinate $\rho$ used above the condition becomes even simpler. From equations \eqref{eq:rotUVf} and \eqref{eq:unrotfg}, we see that in the UV $f$ is an exponential function of $\rho$, so that $f'(\rho)\sim f(\rho)$, and the relevant condition is
\begin{align}
	\frac{g(\rho)}{f(\rho)}=\text{constant} 		\qquad\qquad \text{for $\rho\to\infty$}.
\end{align}
It is clear that the constant-separation behaviour results from a precise cancellation between the functions $f$ and $g$. This can be viewed as the fact that the `exponential' coordinate $\rho$ becomes identical (up to a constant) to the rescaled coordinate $R$ introduced in \autoref{sec:rescaleToR}.

As taking $\rhmin\to\infty$ does not result in $L(\rhmin)\to0$, we find that the UV of the field theory (small separations) is no longer described by the large-$\rho$ region of the bulk theory. This is in contrast to the normal behaviour, and in particular that in the unflavoured case, where increasing $\rhmin$ corresponds to decreasing $L$.

Instead, for separations less than some critical value the `free' solution is stable, as is the case for large separations. In particular $L=0$ corresponds to a string reaching straight down to $\rho=0$, which can be considered a degenerate case of both the `free' and smooth solutions. Aside from this, the smooth solution describes at most only a small range of separations, and in many cases is unstable for all $L$. This describes non-interacting particles.

However, it is not clear how much of this behaviour is physical. In particular, we might expect that resolving the IR singularity would result in $L(\Rb)\ne0$, presumably giving a confining IR as in the unflavoured case (see the `van der Waals' case, \autoref{fig:vanderWaals}). Unless we also then have $L\to0$ for $R\to\Rinf$ we would generically expect a minimum separation, which would be difficult to understand physically.